\documentclass[twocolumn]{aastex62}

\usepackage{graphicx}	% Including figure files
\usepackage{amsmath}	% Advanced maths commands

\usepackage{hyperref}
\usepackage{wrapfig}
\usepackage{subfigure}

\newcommand{\lephare}{{\tt{LePhare}}}
\newcommand{\eazy}{{\tt{EAZY}}}
\newcommand{\bagpipes}{{\tt{Bagpipes}}}

\newcommand{\sextractor}{{\tt{SExtractor}}}
\newcommand{\galfit}{{\tt{GALFIT}}}
\newcommand{\webbpsf}{{\tt{WebbPSF}}}
\newcommand{\MUV}{$M_{\mathrm{UV}}$}
\newcommand{\solm}{$\mathrm{M}_\odot$}

\received{October 3, 2022}
\revised{October XX, 2022}
\accepted{October YY, 2022}

\submitjournal{ApJL}

\shorttitle{JWST NGDEEP High-z Galaxies}
\shortauthors{Austin et al. }

\def\solm{M$_{\odot}\,$}

\def\solm{M$_{\odot}\,$}

\def\casgm20{CAS-G-M$_{20}\,$}
\def\m20{M$_{20}\,$}

\begin{document}

\title{A Large Population of Faint $8<z<16$ Galaxies Found in the First JWST NIRCam Observations of the NGDEEP Survey} 
\correspondingauthor{Duncan Austin}
\email{duncan.austin@postgrad.manchester.ac.uk}

\author[0000-0003-0519-9445]{Duncan Austin}
\affil{Jodrell Bank Centre for Astrophysics, University of Manchester, Oxford Road, Manchester M13 9PL, UK}

\author[0000-0003-4875-6272]{Nathan Adams}
\affil{Jodrell Bank Centre for Astrophysics, University of Manchester, Oxford Road, Manchester M13 9PL, UK}

\author[0000-0003-1949-7638]{Christopher J. Conselice}
\affiliation{Jodrell Bank Centre for Astrophysics, University of Manchester, Oxford Road, Manchester M13 9PL, UK}

\author[0000-0002-4130-636X]{Thomas Harvey}
\affil{Jodrell Bank Centre for Astrophysics, University of Manchester, Oxford Road, Manchester M13 9PL, UK}

\author[0000-0003-2000-3420]{Katherine Ormerod}
\affil{Jodrell Bank Centre for Astrophysics, University of Manchester, Oxford Road, Manchester M13 9PL, UK}

\author[0000-0002-9081-2111]{James Trussler}
\affil{Jodrell Bank Centre for Astrophysics, University of Manchester, Oxford Road, Manchester M13 9PL, UK}

\author[0000-0002-3119-9003]{Qiong Li}
\affil{Jodrell Bank Centre for Astrophysics, University of Manchester, Oxford Road, Manchester M13 9PL, UK}

\author[0000-0002-8919-079X]{Leonardo Ferreira}
\affil{Department of Physics \& Astronomy, University of Victoria, Finnerty Road, Victoria, British Columbia, V8P 1A1, Canada}

\author[0000-0001-8460-1564]{Pratika Dayal}
\affil{Kapteyn Astronomical Institute, University of Groningen, P.O. Box 800, 9700 AV Groningen, The Netherlands}




\begin{abstract}
We present an early analysis on the search for high redshift galaxies using the deepest public \emph{JWST} imaging to date, the NGDEEP field.  This data consists of 6-band NIRCam imaging on the Hubble Ultra Deep Field-Par2, covering a total area of 6.3 arcmin$^{2}$. Based on our initial reduction of the first half of this survey, we reach 5$\sigma$ depths up to mag = 29.5--29.9 between $1-5$~\textmu m.   Such depths present an unprecedented opportunity to begin exploring the early Universe with \emph{JWST}. As such, we find high redshift galaxies in this field by examining the spectral energy distribution of these systems and present 18 new $z > 8$ systems identified using two different photometric redshift codes: \lephare\ and \eazy\, combined with other significance criteria.  The highest redshift object in our sample is at $z=15.57^{+0.39}_{-0.38}$, which has a blue beta slope of $\beta=-3.25^{+0.41}_{-0.46}$ and a very low inferred stellar mass of $M_{*} = 10^{7.39}$~\solm\,. We also discover a series of faint, low-mass dwarf galaxies with $M_{*} < 10^{8.5}$~\solm at $z \sim 9$ that have blue colors and UV slopes. The structure of these galaxies is such that they all have very flat surface brightness profiles and small sizes $< 1 \ \mathrm{kpc}$.  We also compare our results to theory, finding no significant disagreement with some CDM based models.The discovery of these objects, most of which are low luminosity and inferred stellar mass, demonstrates the power of probing continuously deeper into the Universe, pointing the way to deeper, or similar depth but wider area, surveys and demonstrate the critical need for \emph{JWST} deep fields to explore this aspect of the early Universe.
\end{abstract}


\vspace{1em}
\section{Introduction} \label{sec:intro}

The \emph{James Webb Space Telescope} (\emph{JWST}) is quickly revolutionizing our view of the distant Universe and our understanding of when and how galaxy formation occurred at the earliest times \citep{Adams2023,Castellano2022,Finkelstein2022b,Naidu2022,Atek2022,Yan2023,Donnan2022}. One of its key capabilities is the ability to find and study high redshift galaxies, perhaps even up to $z = 20$.  These galaxies are extremely faint and the limitations of past ground- and space-based telescopes have meant galaxies at redshifts greater than around 11 have been nigh impossible to observe. With \emph{JWST}'s immense near-infrared sensitivity, we are now able to observe and study such galaxies in unprecedented detail. %It is also the case that we may be seeing an over abundance of these distant galaxies that are difficult to explain in theoretical models \citep[][]{Lovell2023}.

Incredibly as it may seem at writing, to date the publicly available \emph{JWST} data has not yet exceeded the \textit{Hubble Space Telescope} (\emph{HST}) in terms of depth.  Therefore it remains possible, or even likely, that there are many galaxies waiting to be discovered below the typical depths reached to date. \emph{JWST} is designed to observe primarily in the infrared region, whilst \emph{HST} observes primarily in visible and ultraviolet light.  Most of the discoveries of distant galaxies thus far are due to this redder coverage, rather than any exceptional depth. However, \emph{JWST} has a much larger primary mirror than \emph{HST} that is over 2.5 times larger in diameter.  This provides an ability to probe deeper in the Universe than Hubble, and thus far has been an aspect of the parameter space that has not been explored in any detail beyond examples of gravitational lensing \citep[e.g.,][]{Bhatawdekar2019, Hsiao2022, Pascale2022, Diego2022} or the limited publications from ultra-deep GTO programmes  \citep[e.g.][]{Robertson2022,Curtislake2022,PerezGonzalez2023}. However, the NGDEEP project, which contains a deep NIRCam pointing of the one of the Hubble Ultra Deep Field parallel fields, provides the first opportunity to explore the Universe at an intrinsic depth greater than what Hubble has done to date.

There are many reasons for probing the Universe at a deeper depth than we currently have with existing \emph{JWST} programs.  One reason is that based on early \emph{JWST} data it appears that there may indeed be many more galaxies than expected during this epoch of reionization and beyond \citep[][]{Lovell2023}.   These early results demonstrate that we are finding candidate galaxies upwards of $z>12$ \citep{Adams2023,Castellano2022,Naidu2022,Atek2022,Yan2023,Donnan2022}.  Some of these galaxies have possible confirmed spectroscopic redshifts \citep[][]{Curtislake2022, Fujimoto2023} using NIRSpec observations.  Although no firm conclusions regarding this are available, it is clear that more data is required to address this problem.  As part of the parallel observations of the NGDEEP program, whose primary target is NIRISS spectroscopy of the Hubble Ultra Deep Field (UDF), one of the parallel fields of the UDF (Par-2) has been observed longer and deeper than any public field to date with NIRCam. As such, this gives an excellent opportunity to probe the Universe deeper than we have been able to do to date with \emph{JWST}. 

In this paper, we present the results of these new observations of high redshift galaxies using this deepest data to date taken with \emph{JWST} as part of the NGDEEP observations.   Based on this deep NIRCam imaging, we have discovered 18 high-z galaxies from $8 < z < 16$ and we present in this paper an  examination of their properties in some detail.  These properties include, beyond the  discovery and measured redshifts of these galaxies, their stellar masses, UV slopes as well as their star formation rates.   We discuss how these quantities are measured and compare with previous JWST results for shallower fields. We find that these observations and follow up ones of similar deep fields are revealing new insights into the formation and evolution of galaxies at the very earliest times.   These observations are thus a key aspect towards understanding how galaxy formation progressed, the first time galaxies and stars formed and initial aspects which drive the onset of star formation. We also discuss the implications of our findings for current theories of galaxy formation and evolution and what role they may play in reionizing the Universe.

%Our study also shows that high redshift galaxies are forming stars at a much higher %rate than nearby galaxies, with some forming stars at a rate of up to 1000 solar %masses per year. This is an important discovery, as it helps to explain how the %universe went from being mostly dark and empty to being filled with stars and galaxies.

%The JWST's ability to study high redshift galaxies has also allowed us to study the %properties of the first galaxies, which are thought to have formed just a few hundred %million years after the Big Bang. We found that these galaxies are much smaller than %nearby galaxies, but also much more active, which is consistent with current theories %of galaxy formation.

The structure of the remainder of this paper is as follows.  In Section~\ref{sec:data}, we describe the NGDEEP observations and our reduction, including problems we faced with this unique data set due to its depth. We also describe the data products derived from this new data set which we have created. In Section~\ref{sec:method} we describe our selection procedure undertaken to define a robust sample of galaxies with redshifts $z > 8$. In Section 3 we present an analysis of  the properties of the galaxies we have found. We discuss our results in the context of previous studies and theory in Section 4, and we present a summary of our findings in Section \ref{sec:conclusions}. Throughout this work, we assume a standard cosmology with $H_0=70$\,km\,s$^{-1}$\,Mpc$^{-1}$, $\Omega_{\rm M}=0.3$ and $\Omega_{\Lambda} = 0.7$ to allow for ease of comparison with other observational studies. All magnitudes listed follow the AB magnitude system \citep{Oke1974,Oke1983}.

\section{Data}
\label{sec:data}
%Full colour image with the position of our candidate galaxies circled along with their respective redshifts as a label and colour map.

The Next Generation Deep Extragalactic Exploratory Public (NGDEEP\footnote{DOI: \url{http://dx.doi.org/10.17909/v7ke-ze45}}, PID: 2079, PIs: S.\@ Finkelstein, Papovich and Pirzkal) Survey is a public ultra-deep field which was planned for observations in late-January/early-February of 2023. Papers from the NGDEEP team themselves include Bagley et al. (2023, in prep) for the survey parameters and G. Leung et al. (2023, in prep) for the NIRCam data description.  The primary observation consists of NIRISS Wide Field Slitless Spectroscopy of galaxies within the Hubble Ultra-Deep Field.  Due to a temporary observation suspension of NIRISS during the observation window of the survey, only 50 per cent of observations were taken, with the 2nd half expected in early 2024. Even with this limitation, NGDEEP's NIRCam data in the Hubble Ultra-Deep Field Parallel 2 (HUDF-Par2) is the single deepest public NIRCam observation undertaken in the first 12 months of \emph{JWST}'s operations. These NIRCam observations consist of 6 wide-band NIRCam photometry in 3 short wavelength (SW; F115W, F150W, F200W) and 3 long wavelength (LW; F277W, F356W, F444W) filters. These were taken over 98~ks (F115W), 93~ks (F444W), and 30--42~ks (F150W, F200W, F277W and F356W), of exposure from a combination of SHALLOW4 and DEEP8 readout patterns.  Below we describe the data reduction procedures we use as well as the method for finding the high redshift galaxies in this field.

%There are also NIRISS SW exposures at 1.15, 1.5 and 2.0~microns.
%NIRISS is of HUDF proper, not Par2

\subsection{Data Reduction Process}

We use our own data reduction pipeline first presented in \citet[][]{Adams2023}. Our process consists of running the standard \emph{JWST} pipeline with some minor modifications (pipeline version 1.8.2 and calibration pmap 0995). Between Stages 1 and 2 of the pipeline we apply a correction for 1/F noise and subtract templates of artefacts known as `wisps' from the F150W and F200W imaging. For Stage 2, background subtraction is turned off and replaced with our own 2-dimensional subtraction using {\tt photutils} \citep{Bradley2022}. The images are aligned by using 11 objects that lie within the NIRCam footprint cross matched to the GAIA DR3 database \citep{GAIADR3}. We note the number of GAIA stars in the field is small, therefore further refinement of the WCS is likely required (e.g. by comparing with wider field \emph{HST} or ground-based data) when considering potential follow-up of these sources with precise instruments like NIRSpec. To ensure all NIRCam imaging is self aligned, we further tweak the WCS of the images to align the brightest 200 objects in the field, using the F444W band as the baseline for this.

\begin{figure*}
  
 \includegraphics[width=1\textwidth]{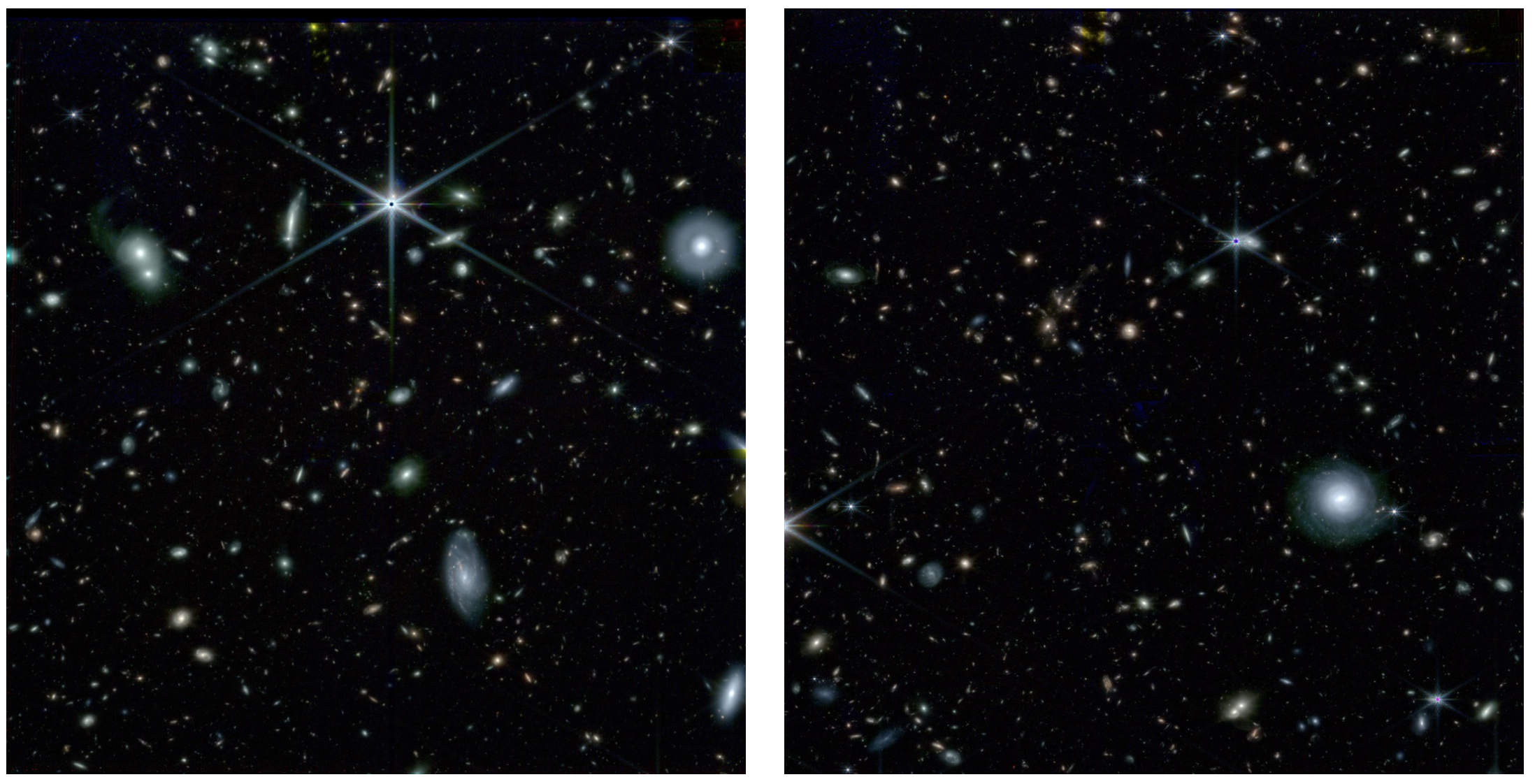}
    
    \caption{An RGB composite of the NGDEEP field after our reduction (R:F444W, G:F277W, B:F200W). Before we carry out our analysis we mask the stars and other bright, foreground galaxies to obtain accurate depths and remove potential spurious detections from our catalogues. }
    \label{fig:rgb}
\end{figure*}

The NGDEEP observations are split into three visits, we find that Visit 3 has a WCS error of approximately 0.7 arcseconds. This resulted in a F200W image offset from the others by 0.7 arcseconds, and a F356W image which containing duplicate objects because its observations were split over multiple visits (such a WCS fault has previously been reported in PRIMER visit 20 of the COSMOS-2 field). We subsequently process Visit 1 and 2 images of F356W together and Visit 3 separately. We then correct the Visit 3 WCS to match the combined Visit 1 and 2 images before stacking these two mosaics together to form the final image (see \autoref{fig:rgb}). For the F115W filter, the data volume is large and the final stage of the \emph{JWST} pipeline struggled to process it. We subsequently split this field into three, equal sized chunks for processing and stacked the final results at the end. 

Due to the depths reached in our reduction, the F150W and F200W filters appear to be limited by the quality of wisp artefact templates that are available. Through experimentation, we discovered that using the initial wisp templates generated by STScI staff in mid 2022 resulted in the affected band having an elevated background noise originating from the templates themselves. This left these modules (particularly module A4 and B3) up to 1 magnitude shallower than the modules that do not require a wisp correction. Upgrading the wisp templates to those released at the end of 2022 resulted in a 0.15-0.2 magnitude improvement in the depths, but indications are that these bands are still limited by the wisp removal process and further fine tuning will be needed in the future.

\subsubsection{Source Extraction}

To locate galaxies we then run \sextractor\ \citep{Bertin1996} with the parameters described in \citet[][]{Adams2023} to obtain forced photometry using F444W as the selection band. We use 0.16 arcsecond radius circular apertures and correct this using models of the PSF from \webbpsf\ \citep{Perrin2014}. This both allows us to observe any potential high-z galaxies whilst retaining faint, lower redshift, blue galaxies due to the low signal requirement of \sextractor\ to extract the source. After the images have passed through our full \emph{JWST} reduction pipeline, we manually mask the shallower outer edges of the image (approx 150 pixels deep), the NIRCam detector gap in the F115W images as well as prominent stellar diffraction spikes and large foreground extended sources which may introduce contaminant flux or false detections. We find that the total unmasked area is $6.32~\mathrm{arcmin}^2$.

\subsubsection{Depth Calculation}

We calculate local depths by placing empty 0.32~arcsec diameter apertures at an approximately constant density in the blank regions of sky in each band, as determined by both our image mask and \sextractor\ segmentation map. Taking the closest 200 apertures to each source in our \sextractor\ catalogue, we re-calculate our photometric errors as the normalized median absolute deviation (NMAD) of the aperture fluxes to include the correlated noise between image pixels, leaving a minimum 5\%\ error to account for future potential zero-point (ZP) issues and other biases. We calculate the average depth over sub-regions of the NGDEEP field by averaging the local depths measured for sources within those regions. A breakdown of these measurements is presented in \autoref{tab:depths}.

\begin{figure}
    \centering
\hspace{-0.5cm}
    \includegraphics[width=0.495\textwidth]{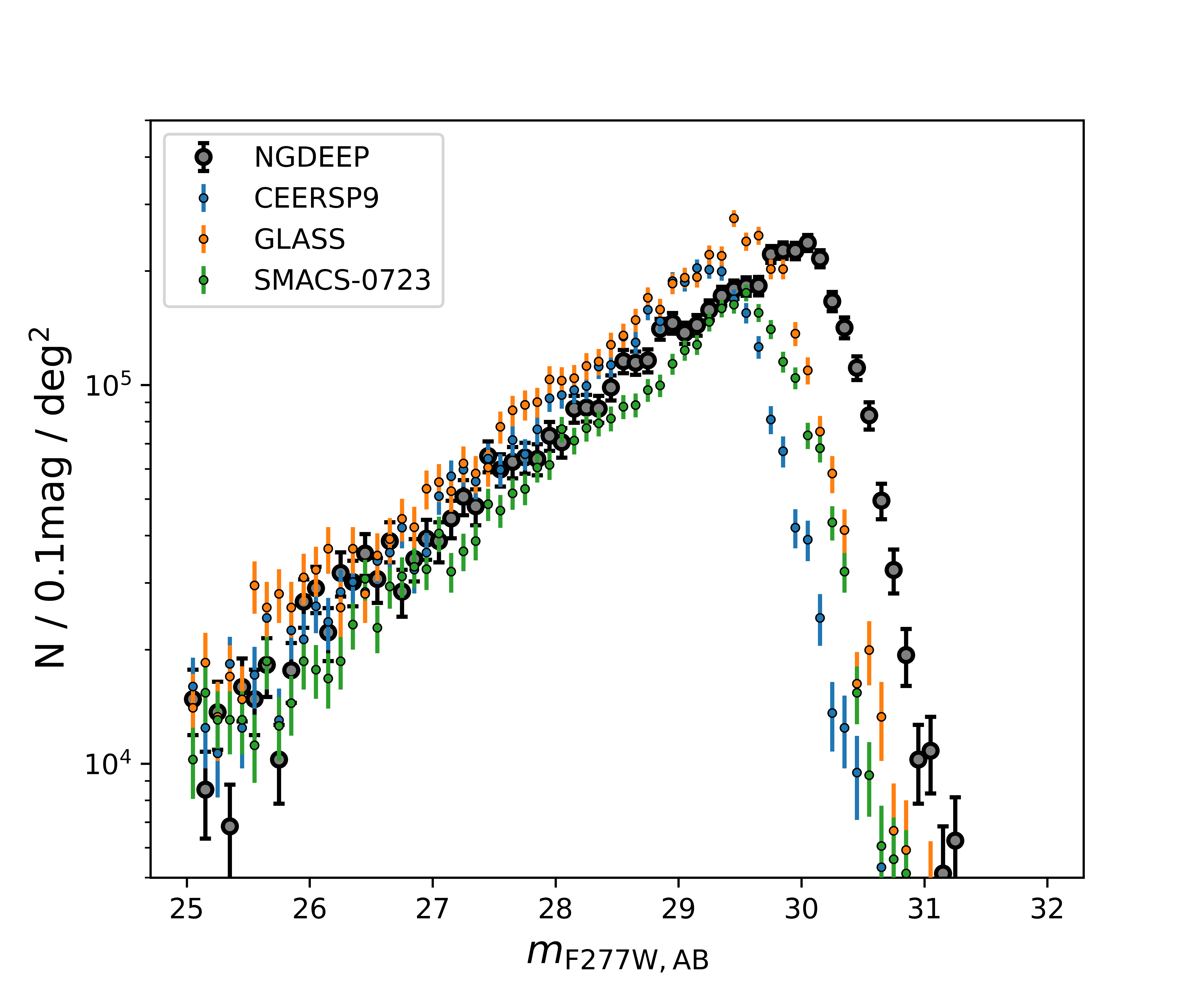}
    \caption{Source counts within the NGDEEP F277W filter in comparison to a selection of other public imaging reduced following the same pipeline. These include SMACS-0723, GLASS and the deepest pointing (P9) of CEERS. We observe the expected result that NGDEEP is around 0.5~mag deeper than GLASS (the previous deepest public survey) which has $5\sigma$ depth of 29.15 in F277W. Displayed errors are purely Poisson and we do not consider the contribution of cosmic variance in this plot.}
    \label{fig:ncounts}
\end{figure}

To estimate how much deeper our data is compared to previous surveys we have compared in Figure~\ref{fig:ncounts} the number counts in the F277W band with the GLASS, CEERS and SMACS 0723 fields.  As can be seen, there are some differences in the number counts between these different fields, in part due to cosmic variance.  However, it can also be seen that the NGDEEP field is half a magnitude deeper in these number counts than in these previously released data.  

\begin{table*}
    \centering
    \caption{Mean $5\sigma$ depths calculated in 0.32~arcsec diameter apertures placed in empty regions of the unmasked area of the GO NGDEEP pointing. We show the depths broken down by sub-groups of NIRCam modules. We find that the central four F150W and F200W modules are significantly shallower than the outer modules. This is likely due to the need for a more precise wisp removal technique. For the red bands, we find the upper center region of module A (spanning a large region around the very luminous star) to be slightly shallower.} %Also shown are $5\sigma$ depths as calculated by the Exposure Time Calculator (ETC) from the exposure times in each band.}
    \label{tab:depths}
\begin{tabular}{ccccccc}
    \hline
    NIRCam filter & F115W & F150W & F200W & F277W & F356W & F444W \\
    \hline
    Outer Modules & 29.65 & 29.75 & 29.65 & 29.80 & 29.75 & 29.60 \\
    Inner Modules & 29.65 & 29.25 & 29.30 & 29.80 & 29.75 & 29.55 \\
    Module A & 29.65 & 29.55 & 29.45 & 29.70 & 29.70 & 29.50 \\
    Module B & 29.65 & 29.50 & 29.55 & 29.90 & 29.80 & 29.60 \\
    Average & 29.65 & 29.50 & 29.50 & 29.80 & 29.75 & 29.55 \\
    \hline
\end{tabular}
\end{table*}

\subsection{Photometric redshifts}

We use the \lephare\ and \eazy\ SED fitting codes to determine redshifts, as well as the size of the Lyman-break as well the significance of detections in various bands. We outline the set-up of \lephare\ and \eazy\ below.

\subsubsection{LePhare}

To calculate preliminary photo-zs, we run our photometric catalogue with updated local-depth errors through the \lephare\ SED fitting code \citep{Arnouts1999,Ilbert2006}. We use the BC03 \citep{Bruzual2003} stellar population synthesis (SPS) template set with both exponentially decaying and constant star formation histories (SFHs) with 10 characteristic timescales between $0.1<\tau<30~\mathrm{Gyr}$, and 57 different ages between 0 and~13 Gyr, with fixed metallicities $Z=\{0.2,~1.0\}~\mathrm{Z}_{\odot}$. The redshift range allowed is $0<z<25$, and we apply dust extinction to these templates up to $E(B-V)<3.5$ in order account for potential dusty lower-z contaminants \citep[e.g.][]{Naidu2022b,Zavala2023}. Attenuation from the inter-galactic medium (IGM) follows the treatment derived in \citep{Madau1995}. \lephare's emission line treatment is also turned on.

\subsubsection{EAZY}

We use a a second SED-fitting tool, \eazy\ \citep{Brammer2008} to confirm our photometric redshifts. We use the default Flexible Stellar Population Synthesis (FSPS) \citep{conroy2010fsps} templates (tweak\_fsps\_QSF\_12\_v3), along with 6 additional templates from \citet{Larson2022}. These templates have been shown to better reproduce the blue colors and $\beta$ slopes of high-z galaxies. The FSPS templates also include a better treatment of emission lines than the BC03 templates, as some high-z galaxies have been shown to have high equivalent width (EW) emission lines, which can boost photometric measurements by as much as a magnitude. 

\subsection{Sample selection}
\label{sec:method}
Based on these photo-zs we  select galaxies using a tiered system to determine ``robust'' and ``good'' galaxy candidates. The criteria for inclusion in these samples are: (1) The galaxy must be $5\sigma$ detected in the 2 bands immediately redward of the inferred Lyman-break and less than $3\sigma$ detected in the bands bluewards of the Lyman break. (2) The integrated probability density function across the primary peak must include more than 60\% percent of the total probability, integrated over $\pm$10\% of the photometric redshift. (3) Any secondary, low-redshift solution, must have a peak probability $<50\%$ of the primary solution. (4) The primary fit must have a $\chi^2_{\mathrm{red}} < 3 (6)$ to be considered robust (good). (5) The above criteria are cross-checked with the results using the second photo-z code \eazy.

In addition, we remove any potential hot pixels from our sample by comparing the \sextractor\ FLUX\_RADIUS parameter to simulated \webbpsf\ \citep{Perrin2014} PSFs in each band, removing sources that are considerably smaller than the NIRCam PSF FWHMs. Such artefacts in the red NIRCam modules can mimic $z=16-20$ photometry. A summary of our NGDEEP galaxy samples, including photometry, photo-z's and galaxy properties is shown in \autoref{tab:sample}. 

\begin{figure*}
     \begin{subfigure}%[b]%{0.49\textwidth}
         \centering
         \includegraphics[width=0.49\textwidth]{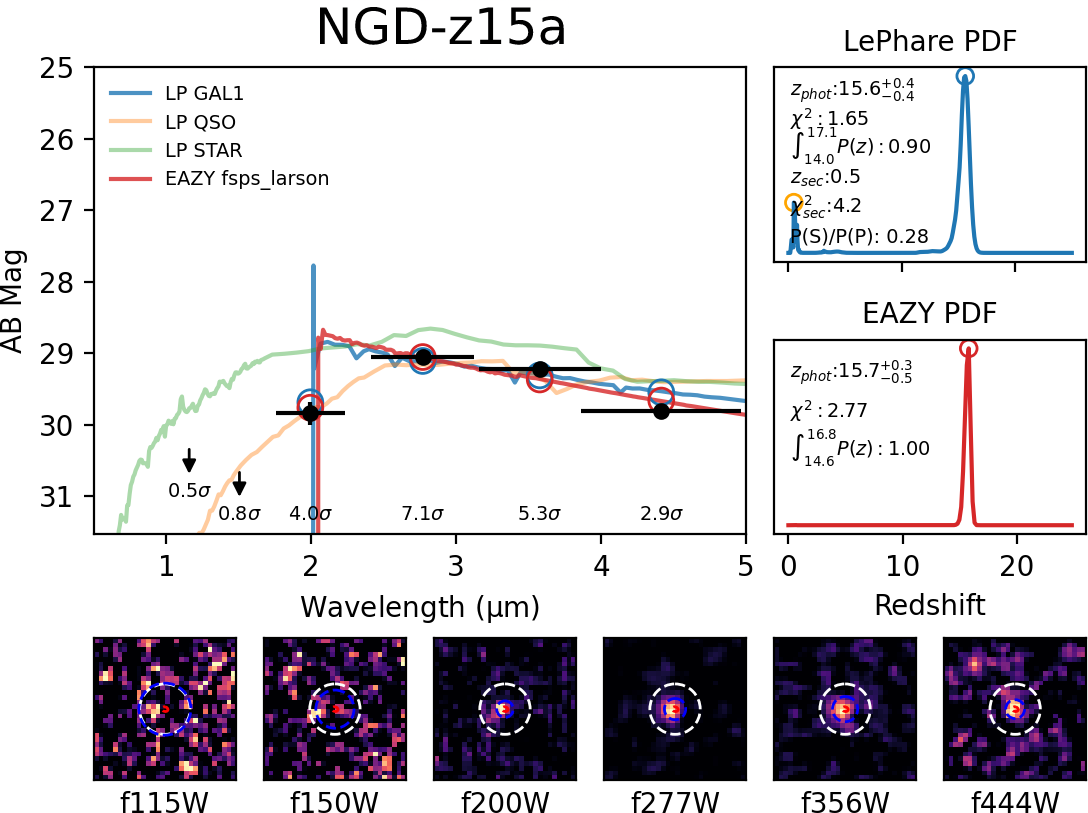}
     \end{subfigure}
     \hfill
     \begin{subfigure}%[b]%{0.49\textwidth}
         \centering
         \includegraphics[width=0.49\textwidth]{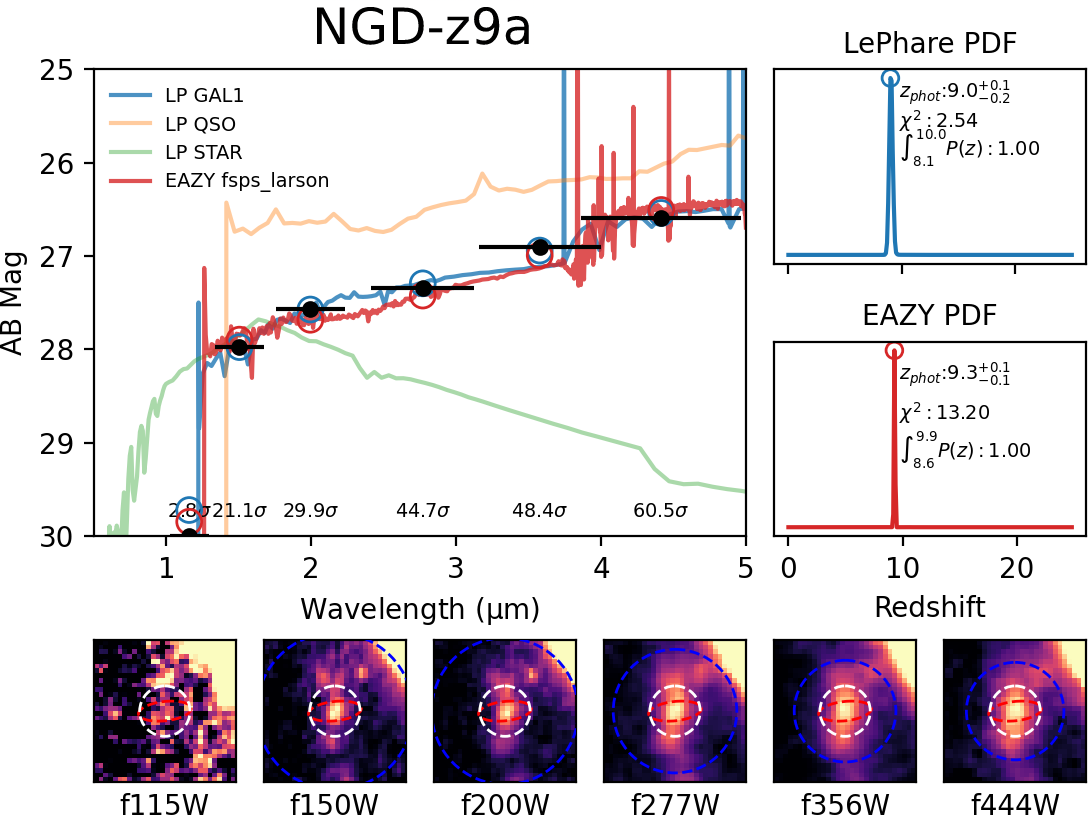}
     \end{subfigure}
     \hfill
     \begin{subfigure}%[b]%{0.49\textwidth}
         \centering
         \includegraphics[width=0.49\textwidth]{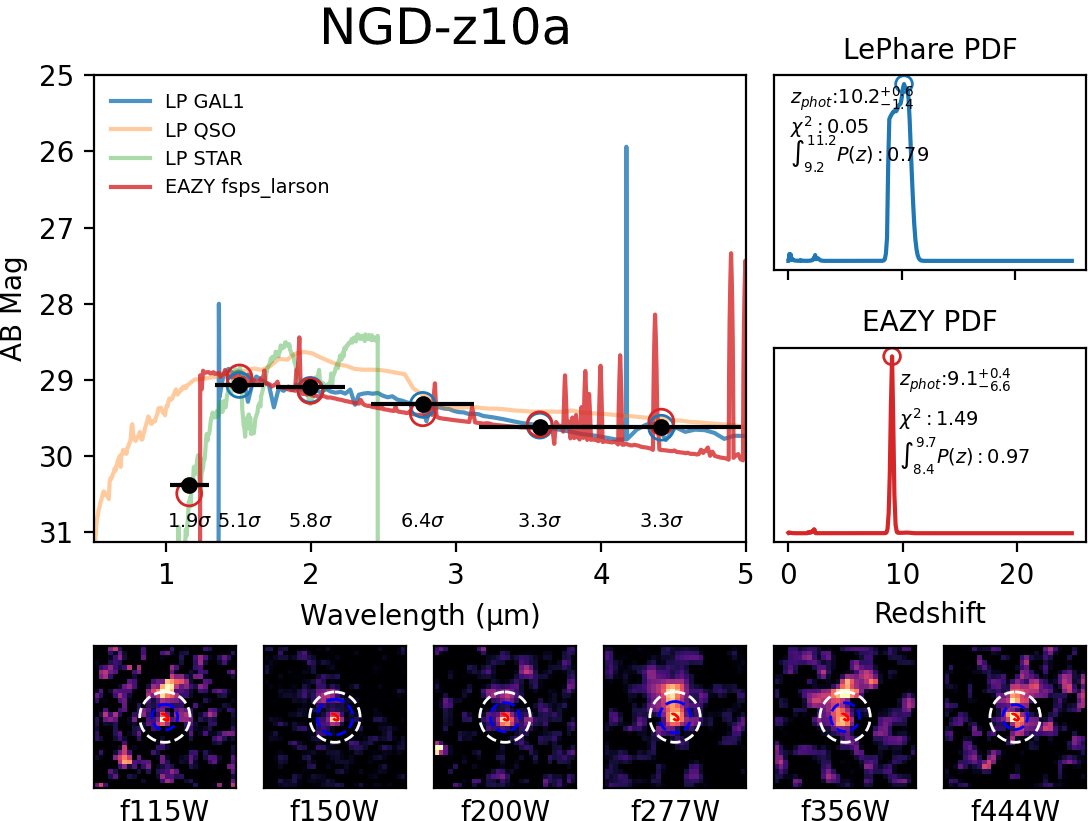}
     \end{subfigure}
    \hfill
    \begin{subfigure}%[b]%{0.49\textwidth}
         \centering
         \includegraphics[width=0.49\textwidth]{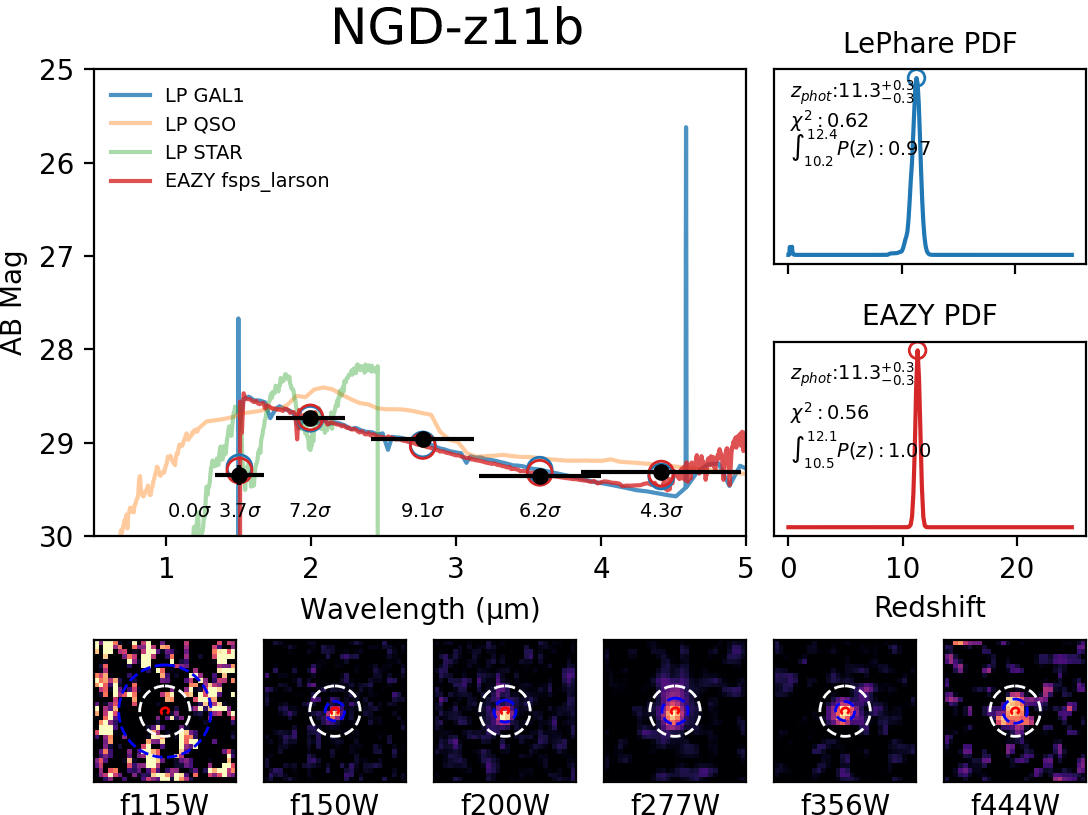}
     \end{subfigure}
     \centering
     \caption{\lephare/\eazy\ SEDs for a selection of galaxies from our sample. The top row shows our highest redshift candidate (left) and the most massive and highest star forming candidate in our sample (right), and the bottom row shows two low-mass dwarf galaxies. On the SED plots the coloured lines show the best fitting \lephare\ and \eazy\ galaxy SEDs, as well as a secondary solution SED and best-fitting brown dwarf and QSO solutions for \lephare\, against our local-depth corrected photometry in black. \lephare/\eazy\ PDFs are shown on the right hand side of each plot, with $\chi^2_{\mathrm{red}}$ shown for the best fitting curves. PDF constraints and secondary solutions are also shown here. $30\times30$ pixel cutouts in each of the 6 wide-band NIRCam filters are shown at the bottom, with \sextractor\ FLUX\_RADIUS shown in dashed blue and our 0.32~arcsec diameter apertures in white.}
\label{fig:exampleseds}
\end{figure*}

\begin{figure}
    \centering
    \includegraphics[width=0.495\textwidth]{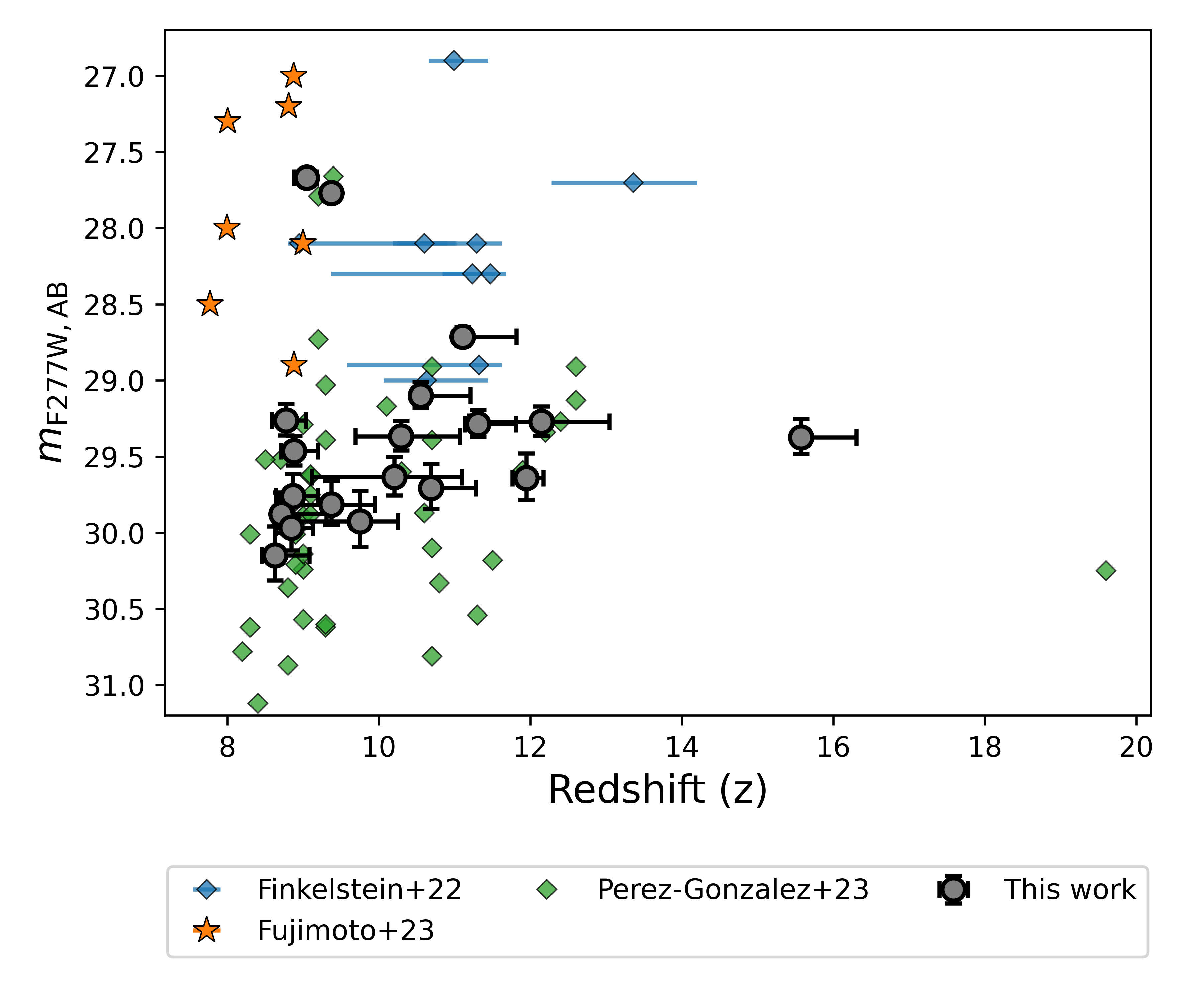}
    \caption{Comparison of our F277W \sextractor\ magnitudes showing the large number of faintly observed candidates we obtain. CEERS spectroscopic \citep[yellow stars]{Fujimoto2023} and photometric \citep[blue diamonds]{Finkelstein2022b} results are shown alongside recent deep data from the semi-overlapping MIRI-DS \citep{PerezGonzalez2023}.}  
    \label{fig:F277W}
\end{figure}

Template fits, images in different bands,  and the probability distributions of redshifts are shown in \autoref{fig:exampleseds}\footnote{All SED plots can be found here \url{https://1drv.ms/u/s!AjXt-wkeMSXAgq52YO_LqY5nvB1NWA?e=KCy4Vr}}.  Likewise we show the distribution of F277W magnitudes for our sample in \autoref{fig:F277W}.  What we find is that many of our new galaxies are fainter than other previous systems. This is already revealing that these systems are of a different nature than the brighter galaxies seen in previous surveys.

%EAZY/LePhare galaxy SEDs and cut outs of our sources here (if few sources), or in appendix if many galaxies.

Due to the lack of F090W in the NGDEEP photometry, and without the inclusion of deep \emph{HST} Advanced Camera for Surveys (ACS) data at 0.6 and 0.8 microns, this work is limited to galaxies at $z>8.5$. The future inclusion of the deep \emph{HST} ACS data will enable for lower redshift sources to be studied in greater detail, including faint $z=4-5$ sources whose Balmer breaks may mimic Lyman breaks in shallower \emph{JWST} fields.

\subsubsection{Comparisons to early HUDF work}

We compare our output catalogues with the $z=8$ sources identified in the work of \citet{Bouwens2011}. While these sources do not meet our strict selection criteria (due to the Lyman break being located within the F115W band), NIRCam observations may be able to validate the \emph{HST}-based redshift estimations for sources that overlap. We cross match all sources within a search radius of 1~arcsecond and examine those with F150W magnitudes close to that of the original \emph{HST} $H_{160}$ band measurements. We obtain 6 total successful cross-matches where redshifts agree. Of these, UDF092y-03781204 has a NIRCam redshift of $z=8.15$, UDF092y-03751196 has a redshift of $z=7.30$ , UDF092y-03391003 has a redshift of $z=8.37$, UDF092y-04242094 has a redshift $z=8.99$, UDF092y-06391247 has a redshift of $z=8.50$, UDF092y-03811034 has a redshfit of $z=8.30$. There are further two sources which lie on the border of our masks and have NIRCam redshifts which do not agree, though this likely due to contamination and breaks too mild in F115W, leading to degeneracies with lower-z Balmer break solutions.

\subsubsection{Comparisons to The MIRI Deep Survey}

The NIRCam NGDEEP observations are not the first to be conducted in the HUDF-Par2 field. The MIRI Deep Survey (PID: 1283, PI's Hans Ulrik Nørgaard-Nielsen, G{\"o}ran {\"O}stlin) has conducted a 4 band NIRCam survey to depths just deeper than magnitude 30. They have reported 45 candidate high-z galaxies \citep{PerezGonzalez2023}, of which around 18 of these are found to have counterparts in our catalogues (this is due to the partial overlap of NGDEEP and MIRI-DS footprints). We find only one of these galaxies enters our final sample, largely due to the low luminosities of the \citet{PerezGonzalez2023} sample failing our 5$\sigma$ selection criteria with very poor SED fits. The brightest source in the overlap region (MDS011049, our NGD-z11a) is successfully recovered with a redshift of $z=11.10^{+0.31}_{-0.46}$ compared to the first published redshift of $z=9.4^{+0.1}_{-0.2}$. Our higher redshift estimate is likely due to the addition of the F200W band in our study, which has a bright measured flux relative to the F150W band, indicating the Lyman break is located partially through the F150W band and not bluewards of the band. We also find that we measure what may be a Balmer break or extreme [O {\sc iii}]/H$\beta$ emission for four faint, cross-matched galaxies in the F444W band, providing redshifts that are uncertain but do agree with \citet{PerezGonzalez2023}. These objects are MDS015081, MDS017690, MDS018332 and MDS030229 between redshifts of $8<z<11$. However, none of our highest redshift galaxies are in this MIRI-DS GTO catalog paper.

\begin{table*}
\setlength{\tabcolsep}{3pt}
\centering

\caption{List of our robust and good NGDEEP galaxy candidates. The first column ascribes a simple name to each galaxy. The second and third columns provide the RA and DEC of each galaxy followed by the observed F444W and F277W AB magnitudes. The sixth and seventh columns are the best fitting redshift and stellar mass from \lephare . Errors are the 16$^{\mathrm{th}}$ and 84$^{\mathrm{th}}$ percentiles of the distribution, with SFR and \MUV\ errors taken from our MCMC fitting to the rest-frame photometry. The `*' (`+') refers to a mass (SFR) which has been corrected for the total flux of the galaxy, rather than the aperture flux used for SED-fitting. SFR and $M_{UV}$ are corrected in the rest-frame UV, and the masses are corrected in the rest-frame optical.}
\label{tab:sample}
\begin{tabular}{llllllllll}
\multicolumn{1}{l}{Name} & \multicolumn{1}{c}{RA} & \multicolumn{1}{c}{DEC} & \multicolumn{1}{c}{F444W} & \multicolumn{1}{c}{F277W} & \multicolumn{1}{c}{Redshift} & \multicolumn{1}{c}{\begin{tabular}[c]{@{}l@{}} Mass\\ $\log M_{*}$\end{tabular}} & \multicolumn{1}{c}{$\beta$} & \multicolumn{1}{c}{\begin{tabular}[c]{@{}l@{}}SFR\\ $\mathrm{M}_{\odot} \textrm{yr}^{-1}$\end{tabular}} & \multicolumn{1}{c}{$M_{\textrm{UV}}$} \\
\hline
\multicolumn{1}{l}{NGD-z8a*+} & \multicolumn{1}{c}{53.23465} & \multicolumn{1}{c}{-27.81601} & \multicolumn{1}{c}{$28.80^{+0.10}_{-0.11}$} & \multicolumn{1}{c}{$29.83^{+0.16}_{-0.19}$} & \multicolumn{1}{c}{$8.63^{+0.30}_{-0.32}$} & \multicolumn{1}{c}{8.38} & \multicolumn{1}{c}{$-2.56^{+0.70}_{-0.71}$} & \multicolumn{1}{c}{$0.98^{+0.57}_{-0.13}$} & \multicolumn{1}{c}{$-18.12^{+0.17}_{-0.15}$} \\
\multicolumn{1}{l}{NGD-z8b*+} & \multicolumn{1}{c}{53.24245} & \multicolumn{1}{c}{-27.80109} & \multicolumn{1}{c}{$29.58^{+0.19}_{-0.23}$} & \multicolumn{1}{c}{$29.55^{+0.12}_{-0.14}$} & \multicolumn{1}{c}{$8.71^{+0.29}_{-0.30}$} & \multicolumn{1}{c}{7.16} & \multicolumn{1}{c}{$-2.38^{+0.61}_{-0.62}$} & \multicolumn{1}{c}{$0.99^{+0.91}_{-0.12}$} & \multicolumn{1}{c}{$-18.12^{+0.14}_{-0.12}$} \\
\multicolumn{1}{l}{NGD-z8c} & \multicolumn{1}{c}{53.24886} & \multicolumn{1}{c}{-27.82388} & \multicolumn{1}{c}{$28.62^{+0.10}_{-0.11}$} & \multicolumn{1}{c}{$28.94^{+0.10}_{-0.11}$} & \multicolumn{1}{c}{$8.77^{+0.21}_{-0.23}$} & \multicolumn{1}{c}{8.14} & \multicolumn{1}{c}{$-2.65^{+0.59}_{-0.60}$} & \multicolumn{1}{c}{$1.54^{+0.35}_{-0.18}$} & \multicolumn{1}{c}{$-18.65^{+0.16}_{-0.14}$} \\
\multicolumn{1}{l}{NGD-z8d} & \multicolumn{1}{c}{53.25345} & \multicolumn{1}{c}{-27.79971} & \multicolumn{1}{c}{$29.03^{+0.11}_{-0.12}$} & \multicolumn{1}{c}{$29.64^{+0.15}_{-0.17}$} & \multicolumn{1}{c}{$8.84^{+0.25}_{-0.24}$} & \multicolumn{1}{c}{8.23} & \multicolumn{1}{c}{$-3.41^{+0.65}_{-0.69}$} & \multicolumn{1}{c}{$1.01^{+0.11}_{-0.10}$} & \multicolumn{1}{c}{$-18.26^{+0.12}_{-0.11}$} \\
\multicolumn{1}{l}{NGD-z8e} & \multicolumn{1}{c}{53.22561} & \multicolumn{1}{c}{-27.80824} & \multicolumn{1}{c}{$29.83^{+0.22}_{-0.28}$} & \multicolumn{1}{c}{$29.44^{+0.13}_{-0.15}$} & \multicolumn{1}{c}{$8.86^{+0.30}_{-0.26}$} & \multicolumn{1}{c}{7.24} & \multicolumn{1}{c}{$-3.18^{+0.58}_{-0.61}$} & \multicolumn{1}{c}{$1.05^{+0.12}_{-0.11}$} & \multicolumn{1}{c}{$-18.30^{+0.12}_{-0.11}$} \\
\multicolumn{1}{l}{NGD-z8f*+} & \multicolumn{1}{c}{53.25449} & \multicolumn{1}{c}{-27.82937} & \multicolumn{1}{c}{$29.05^{+0.12}_{-0.13}$} & \multicolumn{1}{c}{$29.14^{+0.09}_{-0.10}$} & \multicolumn{1}{c}{$8.88^{+0.22}_{-0.27}$} & \multicolumn{1}{c}{7.58} & \multicolumn{1}{c}{$-1.39^{+0.75}_{-0.76}$} & \multicolumn{1}{c}{$5.32^{+13.56}_{-3.78}$} & \multicolumn{1}{c}{$-18.38^{+0.20}_{-0.17}$} \\
\multicolumn{1}{l}{NGD-z9a*+} & \multicolumn{1}{c}{53.23223} & \multicolumn{1}{c}{-27.81654} & \multicolumn{1}{c}{$26.60^{+0.05}_{-0.05}$} & \multicolumn{1}{c}{$27.35^{+0.05}_{-0.05}$} & \multicolumn{1}{c}{$9.05^{+0.15}_{-0.16}$} & \multicolumn{1}{c}{9.91} & \multicolumn{1}{c}{$-0.70^{+0.52}_{-0.51}$} & \multicolumn{1}{c}{$209.18^{+306.81}_{-125.61}$} & \multicolumn{1}{c}{$-21.00^{+0.12}_{-0.10}$} \\
\multicolumn{1}{l}{NGD-z9b*+} & \multicolumn{1}{c}{53.26095} & \multicolumn{1}{c}{-27.82008} & \multicolumn{1}{c}{$27.42^{+0.05}_{-0.05}$} & \multicolumn{1}{c}{$27.45^{+0.05}_{-0.05}$} & \multicolumn{1}{c}{$9.37^{+0.06}_{-0.07}$} & \multicolumn{1}{c}{8.19} & \multicolumn{1}{c}{$-1.73^{+0.50}_{-0.53}$} & \multicolumn{1}{c}{$13.23^{+19.93}_{-7.27}$} & \multicolumn{1}{c}{$-20.07^{+0.10}_{-0.09}$} \\
\multicolumn{1}{l}{NGD-z9c*+} & \multicolumn{1}{c}{53.24578} & \multicolumn{1}{c}{-27.80609} & \multicolumn{1}{c}{$29.21^{+0.14}_{-0.17}$} & \multicolumn{1}{c}{$29.60^{+0.17}_{-0.20}$} & \multicolumn{1}{c}{$9.75^{+0.32}_{-1.01}$} & \multicolumn{1}{c}{7.82} & \multicolumn{1}{c}{$-2.95^{+0.32}_{-0.36}$} & \multicolumn{1}{c}{$1.15^{+0.10}_{-0.10}$} & \multicolumn{1}{c}{$-18.40^{+0.10}_{-0.09}$} \\
\multicolumn{1}{l}{NGD-z10a*+} & \multicolumn{1}{c}{53.26388} & \multicolumn{1}{c}{-27.81721} & \multicolumn{1}{c}{$29.62^{+0.20}_{-0.25}$} & \multicolumn{1}{c}{$29.31^{+0.12}_{-0.13}$} & \multicolumn{1}{c}{$10.20^{+0.61}_{-1.37}$} & \multicolumn{1}{c}{7.12} & \multicolumn{1}{c}{$-2.59^{+0.55}_{-0.55}$} & \multicolumn{1}{c}{$1.52^{+0.43}_{-0.25}$} & \multicolumn{1}{c}{$-18.59^{+0.25}_{-0.23}$} \\
\multicolumn{1}{l}{NGD-z10b*+} & \multicolumn{1}{c}{53.24223} & \multicolumn{1}{c}{-27.83051} & \multicolumn{1}{c}{$28.91^{+0.16}_{-0.18}$} & \multicolumn{1}{c}{$29.04^{+0.09}_{-0.10}$} & \multicolumn{1}{c}{$10.29^{+0.72}_{-0.66}$} & \multicolumn{1}{c}{8.12} & \multicolumn{1}{c}{$-1.96^{+0.62}_{-0.53}$} & \multicolumn{1}{c}{$7.85^{+10.65}_{-2.33}$} & \multicolumn{1}{c}{$-19.91^{+0.29}_{-0.25}$} \\
\multicolumn{1}{l}{NGD-z10c} & \multicolumn{1}{c}{53.27685} & \multicolumn{1}{c}{-27.85078} & \multicolumn{1}{c}{$28.68^{+0.08}_{-0.09}$} & \multicolumn{1}{c}{$28.78^{+0.08}_{-0.09}$} & \multicolumn{1}{c}{$10.55^{+0.27}_{-0.31}$} & \multicolumn{1}{c}{8.21} & \multicolumn{1}{c}{$-3.29^{+0.41}_{-0.43}$} & \multicolumn{1}{c}{$2.93^{+0.45}_{-0.40}$} & \multicolumn{1}{c}{$-19.42^{+0.16}_{-0.16}$} \\
\multicolumn{1}{l}{NGD-z10d*+} & \multicolumn{1}{c}{53.25869} & \multicolumn{1}{c}{-27.80494} & \multicolumn{1}{c}{$29.30^{+0.15}_{-0.17}$} & \multicolumn{1}{c}{$29.38^{+0.14}_{-0.16}$} & \multicolumn{1}{c}{$10.69^{+0.33}_{-0.35}$} & \multicolumn{1}{c}{7.98} & \multicolumn{1}{c}{$-2.89^{+0.51}_{-0.55}$} & \multicolumn{1}{c}{$1.92^{+0.31}_{-0.25}$} & \multicolumn{1}{c}{$-18.93^{+0.17}_{-0.16}$} \\
\multicolumn{1}{l}{NGD-z11a*+} & \multicolumn{1}{c}{53.26715} & \multicolumn{1}{c}{-27.84908} & \multicolumn{1}{c}{$28.06^{+0.05}_{-0.05}$} & \multicolumn{1}{c}{$28.39^{+0.06}_{-0.06}$} & \multicolumn{1}{c}{$11.10^{+0.31}_{-0.46}$} & \multicolumn{1}{c}{9.07} & \multicolumn{1}{c}{$-1.61^{+0.42}_{-0.44}$} & \multicolumn{1}{c}{$7.50^{+7.91}_{-3.96}$} & \multicolumn{1}{c}{$-19.21^{+0.15}_{-0.13}$} \\
\multicolumn{1}{l}{NGD-z11b} & \multicolumn{1}{c}{53.24201} & \multicolumn{1}{c}{-27.85526} & \multicolumn{1}{c}{$29.31^{+0.16}_{-0.19}$} & \multicolumn{1}{c}{$28.97^{+0.09}_{-0.09}$} & \multicolumn{1}{c}{$11.31^{+0.32}_{-0.35}$} & \multicolumn{1}{c}{7.35} & \multicolumn{1}{c}{$-2.62^{+0.48}_{-0.45}$} & \multicolumn{1}{c}{$2.01^{+0.36}_{-0.23}$} & \multicolumn{1}{c}{$-18.95^{+0.16}_{-0.15}$} \\
\multicolumn{1}{l}{NGD-z11c*+} & \multicolumn{1}{c}{53.27762} & \multicolumn{1}{c}{-27.86748} & \multicolumn{1}{c}{$29.29^{+0.20}_{-0.25}$} & \multicolumn{1}{c}{$29.32^{+0.14}_{-0.16}$} & \multicolumn{1}{c}{$11.95^{+0.19}_{-0.22}$} & \multicolumn{1}{c}{7.39} & \multicolumn{1}{c}{$-4.64^{+0.51}_{-0.56}$} & \multicolumn{1}{c}{$3.09^{+0.34}_{-0.34}$} & \multicolumn{1}{c}{$-19.48^{+0.13}_{-0.11}$} \\
\multicolumn{1}{l}{NGD-z12a} & \multicolumn{1}{c}{53.26652} & \multicolumn{1}{c}{-27.87676} & \multicolumn{1}{c}{$28.75^{+0.11}_{-0.12}$} & \multicolumn{1}{c}{$28.95^{+0.09}_{-0.10}$} & \multicolumn{1}{c}{$12.15^{+1.28}_{-0.59}$} & \multicolumn{1}{c}{8.94} & \multicolumn{1}{c}{$-2.12^{+0.43}_{-0.46}$} & \multicolumn{1}{c}{$2.13^{+2.37}_{-0.36}$} & \multicolumn{1}{c}{$-18.84^{+0.12}_{-0.11}$} \\
\multicolumn{1}{l}{NGD-z15a} & \multicolumn{1}{c}{53.24942} & \multicolumn{1}{c}{-27.87590} & \multicolumn{1}{c}{$29.81^{+0.23}_{-0.29}$} & \multicolumn{1}{c}{$29.05^{+0.11}_{-0.12}$} & \multicolumn{1}{c}{$15.57^{+0.39}_{-0.38}$} & \multicolumn{1}{c}{7.39} & \multicolumn{1}{c}{$-3.25^{+0.41}_{-0.46}$} & \multicolumn{1}{c}{$2.46^{+0.34}_{-0.32}$} & \multicolumn{1}{c}{$-19.23^{+0.15}_{-0.14}$} \\
\hline
\end{tabular}

\end{table*}

\section{NGDEEP High-z galaxy properties}
\label{sec:Properties}

As we perform our SED fitting using fluxes taken from fixed 0.32 arcsec apertures, we correct the derived masses and SFRs for extended sources using \sextractor 's FLUX\_AUTO, which captures $>95\%$ of the total flux of each object in an elliptical Kron aperture. We calculate mass and SFR correction factors from the ratio of FLUX\_AUTO to our aperture corrected aperture fluxes, using F444W for the mass and the band closest to the rest frame UV at 1500~\AA\ for the SFR and $M_{\mathrm{UV}}$.  This ensures that we are obtaining the total light from these objects regardless of their size or morphology.

\subsection{Stellar masses, SFRs and $\beta$ slopes}

In this section we discuss the stellar masses, star formation rates (SFRs) and the $\beta$ slopes of our sample. These quantities are important for understanding the formation state and the physical mechanisms at play in these galaxies and how they relate to previously discovered populations.  
%In addition, we determine the rest-frame UV properties of the most prominent secondary solutions for our sample. 

%The calculated SFRs are incorrect due to the possibility of an incorrect conversion factor being used. This can only be unscrambled using high S/N NIR spectra from JWSTs NIRSpec. 

%A thorough comparison of SFRs calculated using \bagpipes\ parametric SFHs and this method is presented in Austin et al. 2023 (in prep.).

\begin{figure}
    \centering
    \includegraphics[width=0.495\textwidth]{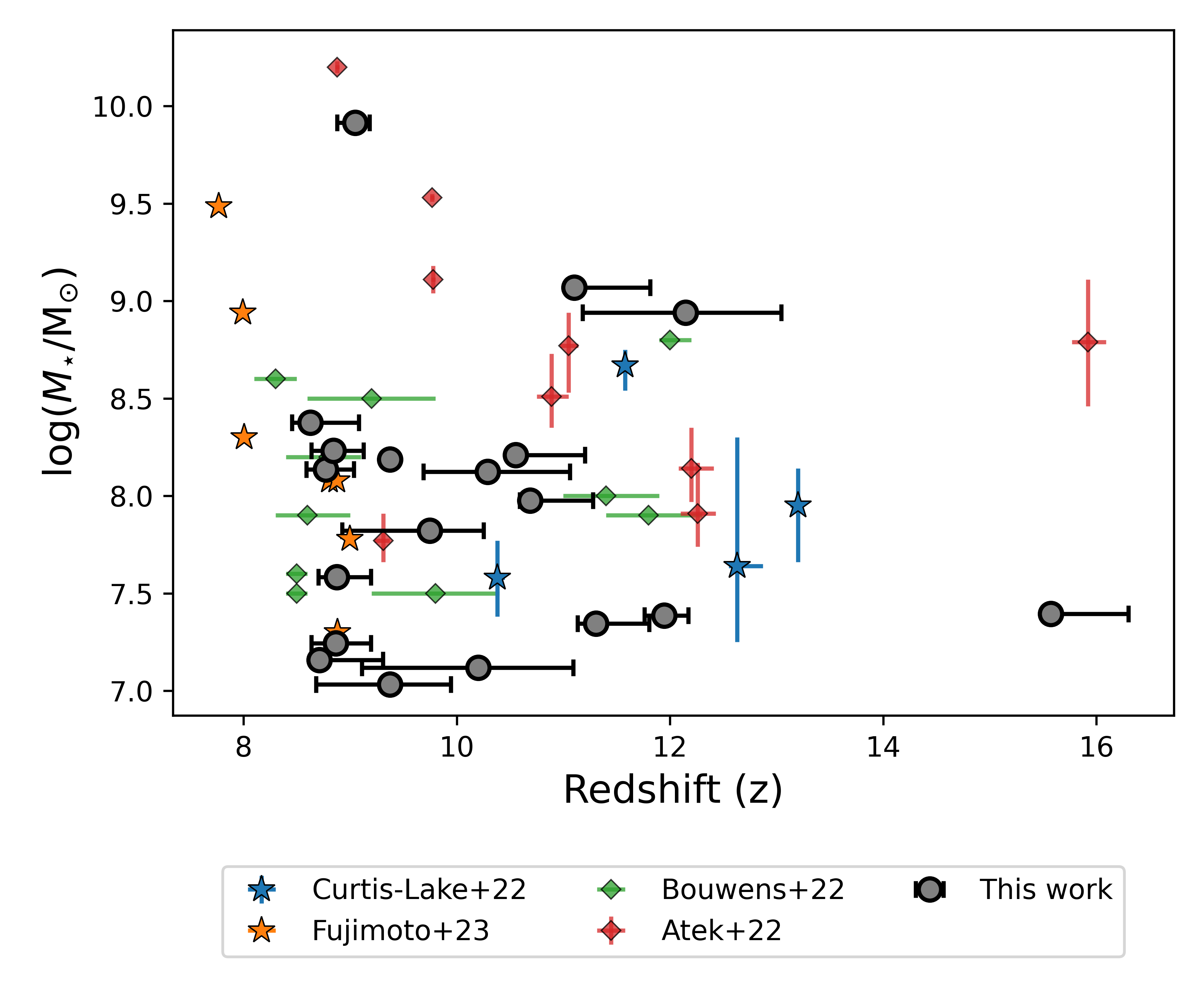}
    \caption{Corrected \lephare\ stellar masses as a function of redshift for our NGDEEP candidates. We find an abundance of low mass galaxies, with $\sim50\%$ of our sample having \lephare\ masses $M_*<10^8$~$\mathrm{M}_{\odot}$.}
    \label{fig:massz}
\end{figure}

We compare our stellar masses as derived from \lephare\, to stellar masses calculated from previous \emph{JWST} work using the GLASS, SMACS 0723 and CEERS fields.  Figure~\ref{fig:massz} shows the distribution of detected stellar masses vs. redshift. These stellar masses are calculated with a standard \citet{Chabrier2003} IMF and thus there could indeed be systematic deviations from true stellar masses in these calculations if the IMF is different at high redshifts. Regardless, this allows us to compare how a standard measured stellar mass compares to these shallower surveys.  What we can see is that with NGDEEP we are finding many more low mass galaxies at $z > 8.5$, which if these masses are accurate puts these systems into the regime of dwarf galaxies. This is a sign that there are many more faint low mass galaxies at these high redshifts to discover in deep \emph{JWST} data. 

\begin{figure}
    \centering
    \includegraphics[width=0.495\textwidth]{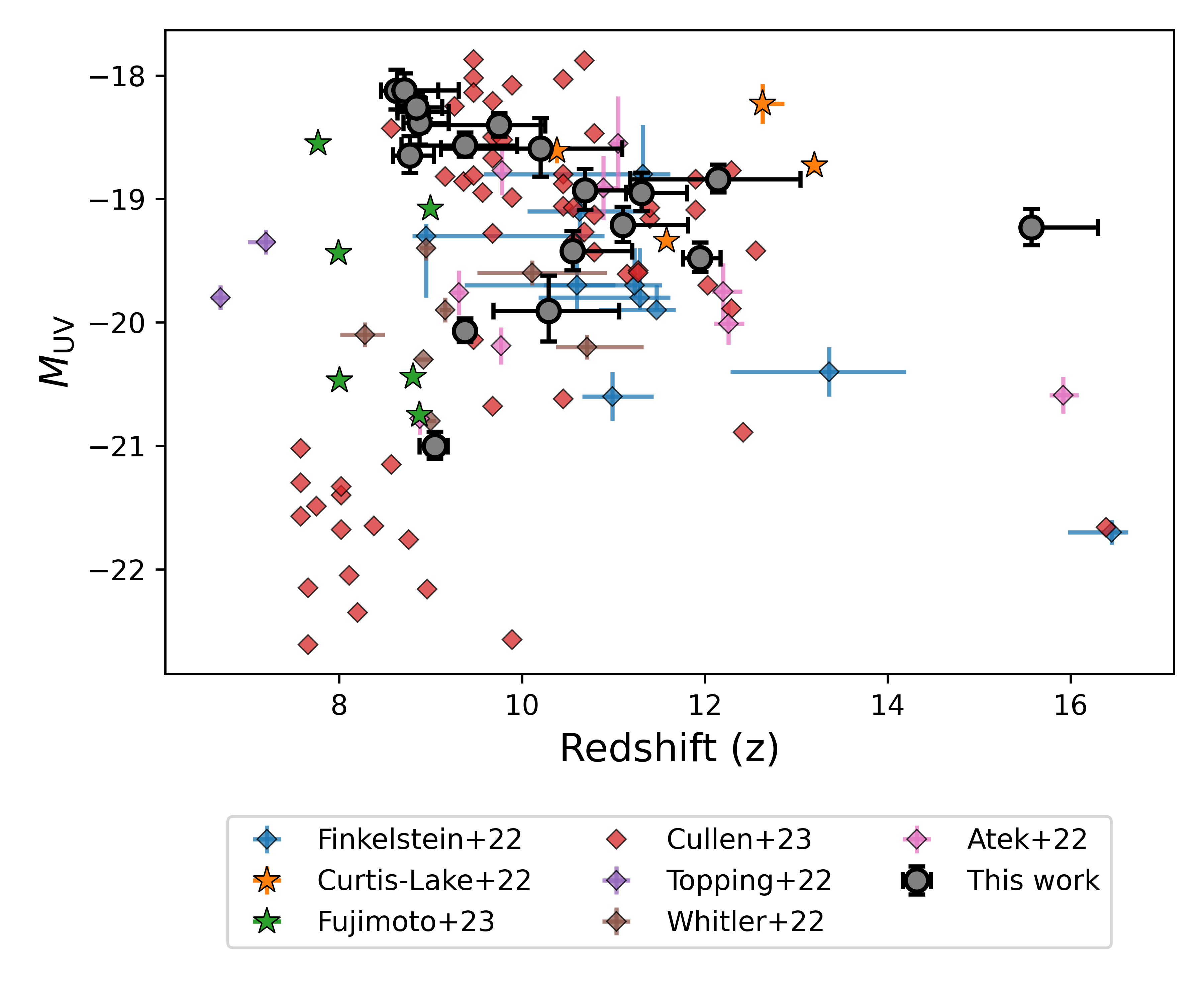}
    \caption{The sample absolute ultraviolet luminosity plotted against redshift. Plotted as stars are spectroscopically confirmed galaxies from JADES \citep{Curtislake2022} and CEERS \citep{Fujimoto2023}. NIRCam selected candidate galaxies from CEERS \citep{Finkelstein2022b} and early \emph{JWST} papers \citep{Cullen2023, Topping2022, Whitler2023, Atek2022} are shown as smaller diamonds. The \citet{Cullen2023} sources with \MUV\ $<-21$ show the parameter space explored by UltraVISTA \citep{McCracken2012}. Our NGD-z15a is unique in that it is the faintest and least massive object at $z>13.5$ seen in \emph{JWST} data.}
    \label{fig:MUVz}
\end{figure}

\begin{figure}
    \centering
    \includegraphics[width=0.495\textwidth]{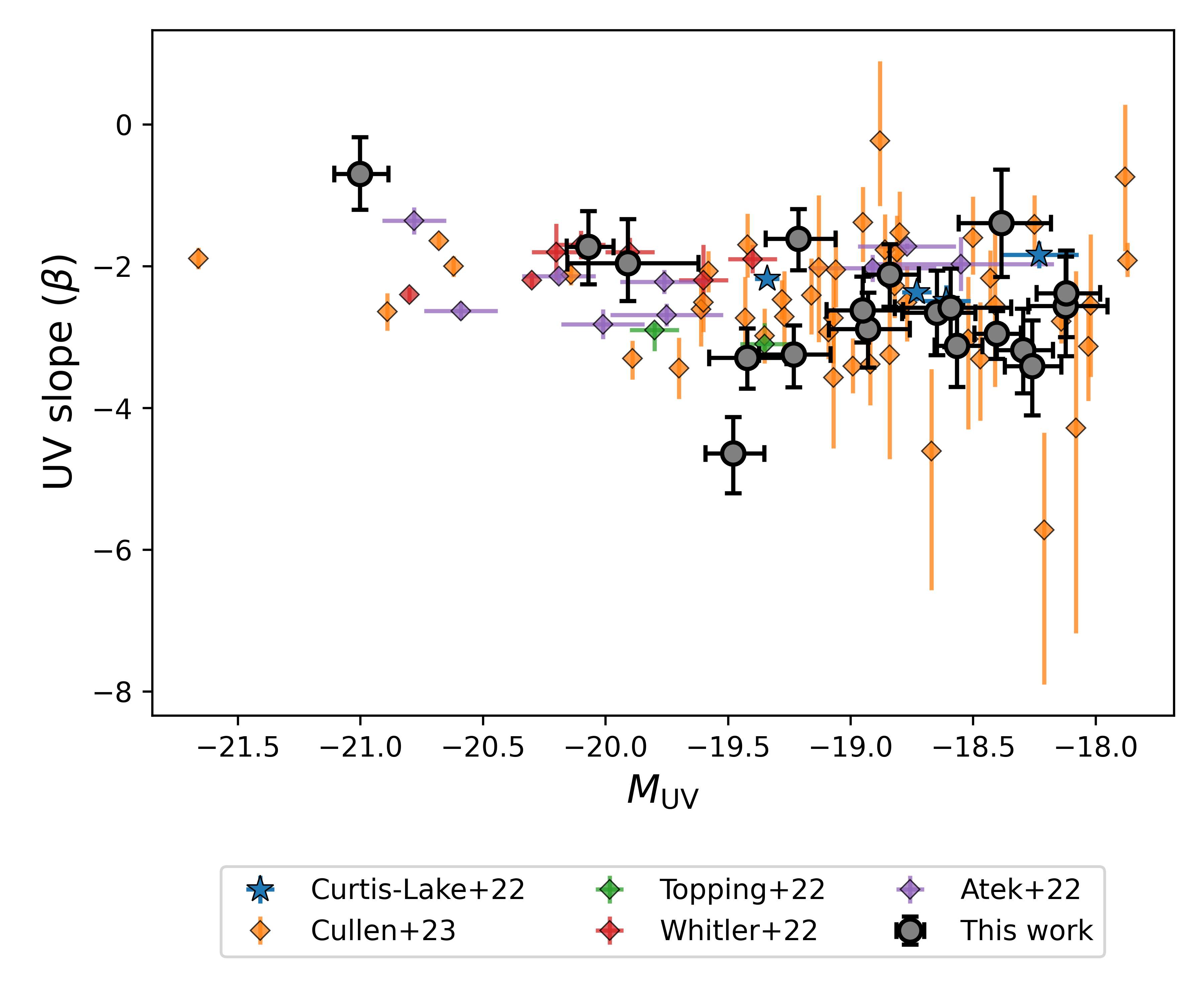}
    \caption{Ultraviolet spectral slope ($\beta$) as a function of absolute ultraviolet magnitude (\MUV) for our NGDEEP candidates. We compare to NIRCam selected candidates from \citet{Cullen2023} (orange), \citet{Topping2022} (green), \citet{Whitler2023} (red) and \citet{Atek2022} (purple). The 4 \citet{Curtislake2022} spectroscopically confirmed galaxies are also shown as blue stars. There is good agreement between the majority of our results and previous literature, except for the extremely blue NGD-z11c and NGD-z9a which is our intrinsically brightest and reddest candidate.}
    \label{fig:betaMUV}
\end{figure}

We follow the procedure in \citet{Bhatawdekar2019} and Austin et al. 2023 (in prep.) to calculate near SED template independent (all but derived redshift) SFRs. We fit the rest-frame UV photometry ($1216<\lambda_{\mathrm{rest}}~/~\AA <3000$) whilst fixing the best-fitting redshift from \lephare\ with a power law of the form $f_{\lambda}\propto\lambda^\beta$. This fit is conducted with the use of {\tt emcee} \citep{emcee} to obtain Bayesian errors. From this, we determine UV continuum slopes, $\beta$, directly and $M_{\mathrm{UV}}$ by averaging the flux within a top-hat of width 100~\AA\ centred on 1500~\AA\, applying the same UV correction factor outlined at the top of \autoref{sec:Properties}. We show a comparison of our calculated $M_{\mathrm{UV}}$ values as a function of $z$ in Figure~\ref{fig:MUVz}. From this rest-frame UV flux, we calculate the observed UV luminosity correcting for dust as per the \citet{Meurer1999} relation and converting to SFR using the \citet{Madau2014} factor. We notice that our calculated $\beta$ slopes appear as expected, with the majority of our candidates having $-3<\beta<-1.5$. Our NGD-z11c is surprisingly blue and far beyond the limit expected from our SED template sets. This could be a result of either photometric uncertainties, as noted in \citet{Cullen2023}, or a system with increased Lyman-alpha continuum emission \citet{Topping2022}. Our $\beta$ slopes are plotted as a function of $M_{\mathrm{UV}}$ in \autoref{fig:betaMUV}. Calculated masses and SFRs are also shown in Table~\ref{tab:sample}.

\subsection{Non-Parametric Mass and SFR Fitting}
\label{sec:Bagpipes}

We use the Bayesian SED fitting code \bagpipes\ \citep[][]{Carnall2018_Bagpipes} to determine galaxy properties such as stellar masses, SFRs, dust content and metallicity for our galaxies in NGDEEP. The latter two properties are less well constrained with photometric data alone, so we do not report those values here. Table~\ref{tab:bagpipes} shows the stellar mass and SFR estimates for these galaxies.

We use the built-in BC03 template set and set logarithmic priors for the age, metallicity and dust whilst fixing the redshift to the best-fitting \lephare\ solution\footnote{Note that fixing our redshift means the errors on our masses/SFRs are underestimated as the photometric redshift PDFs are not taken into account in our analysis.}. We run \bagpipes\ three times with different star formation histories (SFHs; exponential, delayed exponential and constant) whilst assuming a \citet{Calzetti2000} dust extinction law, which has been found to be closely followed at high redshifts (Bowler et al. 2023, in prep.). We allow the stellar mass to be fit between 5  $\leq \log(M_*/M_{\odot})\leq$ 12. The age is allowed to vary between 0 and 15 Gyr, and for the exponential and delayed SFH models the e-folding timescale, $\tau$, is allowed to vary between 0.01 and 15 Gyr. For the dust extinction $A_V$ is fit between 0 and 6 magnitudes, and for nebular emission the ionization parameter ($U$) is fit from $-4 \leq \log U \leq -2$. The metallicity is allowed to vary between -4  $\leq \log(Z_*/\mathrm{Z}_{\odot}) \leq$ 1. We also reduce the star formation timescale from the base \bagpipes\ code to 10~Myr to allow for the expected increase in sSFR for galaxies at these redshifts compared to those at lower-z. 

We carry out these calculations to determine the range of SFR and masses for our sample as compared with the \lephare\ observations which we use in our plots.  We find, roughly independent of the method of the assumed star formation that the stellar masses and star formation rates roughly agree for these early galaxies with our SED fitting results.  We also find, unsurprisingly, that a constant star formation history is unlikely to be an accurate representation of these galaxies.  The exponential fits and delayed exponential fits give similar results, and with our fitting cannot distinguish which is the better of the two. 

\begin{table*}
\setlength{\tabcolsep}{0.1pt}
\caption{Table showing derived stellar masses and SFR using the Bayesian SED-fitting tool Bagpipes. Three different SFHs were used, with `del', `exp', and `const', referring to a delayed exponential model (SFR $\propto t e^{-t/\tau})$, an exponential model (SFR  $\propto e^{-t/\tau})$, and a constant SFR respectively. A `*' or a `+' marks where the stellar mass or SFR has been corrected for total flux of the galaxy outside our photometric apertures. All masses quoted are stellar masses.}
\label{tab:bagpipes}
\centering
\begin{tabular}{lllllll}
\multicolumn{1}{c}{Name} & \multicolumn{1}{c}{\begin{tabular}[c]{@{}l@{}} Mass (del)\end{tabular}} & \multicolumn{1}{c}{\begin{tabular}[c]{@{}l@{}}SFR (del)\end{tabular}} & \multicolumn{1}{c}{\begin{tabular}[c]{@{}l@{}} Mass (exp)\end{tabular}} & \multicolumn{1}{c}{\begin{tabular}[c]{@{}l@{}}SFR (exp) \end{tabular}} & \multicolumn{1}{c}{\begin{tabular}[c]{@{}l@{}} Mass (const)\end{tabular}} & \multicolumn{1}{c}{\begin{tabular}[c]{@{}l@{}}SFR (const)\end{tabular}} \\
\multicolumn{1}{c}{} & \multicolumn{1}{c}{\begin{tabular}[c]{@{}l@{}}$\log_{10}(M_{*}/\mathrm{M}_{\odot})$\end{tabular}} & \multicolumn{1}{c}{\begin{tabular}[c]{@{}l@{}} $\mathrm{M}_{\odot} \textrm{yr}^{-1}$\end{tabular}} & \multicolumn{1}{c}{\begin{tabular}[c]{@{}l@{}}$\log_{10}(M_{*}/\mathrm{M}_{\odot})$\end{tabular}} & \multicolumn{1}{c}{\begin{tabular}[c]{@{}l@{}} $\mathrm{M}_{\odot} \textrm{yr}^{-1}$\end{tabular}} & \multicolumn{1}{c}{\begin{tabular}[c]{@{}l@{}} $\log_{10}(M_{*}/\mathrm{M}_{\odot})$\end{tabular}} & \multicolumn{1}{c}{\begin{tabular}[c]{@{}l@{}} $\mathrm{M}_{\odot} \textrm{yr}^{-1}$\end{tabular}} \\
\hline
\multicolumn{1}{c}{NGD-z8a$^{*+}$} & \multicolumn{1}{c}{$8.33^{+0.34}_{-0.08}$} & \multicolumn{1}{c}{$0.93^{+0.07}_{-0.30}$} & \multicolumn{1}{c}{$8.55^{+0.13}_{-0.17}$} & \multicolumn{1}{c}{$0.74^{+0.22}_{-0.16}$} & \multicolumn{1}{c}{$8.55^{+0.12}_{-0.14}$} & \multicolumn{1}{c}{$0.75^{+0.22}_{-0.15}$} \\
\multicolumn{1}{c}{NGD-z8b$^{*+}$} & \multicolumn{1}{c}{$7.81^{+0.30}_{-0.34}$} & \multicolumn{1}{c}{$0.67^{+0.18}_{-0.13}$} & \multicolumn{1}{c}{$7.98^{+0.18}_{-0.15}$} & \multicolumn{1}{c}{$0.55^{+0.10}_{-0.06}$} & \multicolumn{1}{c}{$7.80^{+0.31}_{-0.33}$} & \multicolumn{1}{c}{$0.63^{+0.21}_{-0.10}$} \\
\multicolumn{1}{c}{NGD-z8c} & \multicolumn{1}{c}{$8.37^{+0.25}_{-0.07}$} & \multicolumn{1}{c}{$1.35^{+0.13}_{-0.49}$} & \multicolumn{1}{c}{$8.50^{+0.13}_{-0.15}$} & \multicolumn{1}{c}{$1.08^{+0.35}_{-0.22}$} & \multicolumn{1}{c}{$8.50^{+0.12}_{-0.20}$} & \multicolumn{1}{c}{$1.10^{+0.38}_{-0.24}$} \\
\multicolumn{1}{c}{NGD-z8d} & \multicolumn{1}{c}{$8.43^{+0.26}_{-0.01}$} & \multicolumn{1}{c}{$0.78^{+0.00}_{-0.27}$} & \multicolumn{1}{c}{$8.59^{+0.11}_{-0.15}$} & \multicolumn{1}{c}{$0.60^{+0.18}_{-0.10}$} & \multicolumn{1}{c}{$8.59^{+0.10}_{-0.15}$} & \multicolumn{1}{c}{$0.60^{+0.18}_{-0.09}$} \\
\multicolumn{1}{c}{NGD-z8e} & \multicolumn{1}{c}{$7.66^{+0.28}_{-0.29}$} & \multicolumn{1}{c}{$0.72^{+0.18}_{-0.14}$} & \multicolumn{1}{c}{$7.97^{+0.13}_{-0.11}$} & \multicolumn{1}{c}{$0.58^{+0.08}_{-0.06}$} & \multicolumn{1}{c}{$7.64^{+0.30}_{-0.27}$} & \multicolumn{1}{c}{$0.71^{+0.19}_{-0.13}$} \\
\multicolumn{1}{c}{NGD-z8f$^{*+}$} & \multicolumn{1}{c}{$8.15^{+0.27}_{-0.27}$} & \multicolumn{1}{c}{$1.19^{+0.28}_{-0.33}$} & \multicolumn{1}{c}{$8.30^{+0.17}_{-0.16}$} & \multicolumn{1}{c}{$0.94^{+0.20}_{-0.14}$} & \multicolumn{1}{c}{$8.19^{+0.23}_{-0.31}$} & \multicolumn{1}{c}{$1.06^{+0.40}_{-0.21}$} \\
\multicolumn{1}{c}{NGD-z9a$^{*+}$} & \multicolumn{1}{c}{$9.84^{+0.21}_{-0.33}$} & \multicolumn{1}{c}{$215.84^{+14.16}_{-23.96}$} & \multicolumn{1}{c}{$10.07^{+0.13}_{-0.13}$} & \multicolumn{1}{c}{$105.06^{+12.99}_{-7.07}$} & \multicolumn{1}{c}{$9.84^{+0.21}_{-0.33}$} & \multicolumn{1}{c}{$211.67^{+15.07}_{-23.06}$} \\
\multicolumn{1}{c}{NGD-z9b$^{*+}$} & \multicolumn{1}{c}{$8.45^{+0.21}_{-0.24}$} & \multicolumn{1}{c}{$8.16^{+1.73}_{-1.78}$} & \multicolumn{1}{c}{$8.84^{+0.08}_{-0.08}$} & \multicolumn{1}{c}{$5.38^{+0.67}_{-0.95}$} & \multicolumn{1}{c}{$8.46^{+0.19}_{-0.25}$} & \multicolumn{1}{c}{$8.37^{+1.52}_{-1.99}$} \\
\multicolumn{1}{c}{NGD-z9c$^{*+}$} & \multicolumn{1}{c}{$8.09^{+0.33}_{-0.03}$} & \multicolumn{1}{c}{$0.81^{+0.06}_{-0.21}$} & \multicolumn{1}{c}{$8.30^{+0.12}_{-0.22}$} & \multicolumn{1}{c}{$0.68^{+0.15}_{-0.09}$} & \multicolumn{1}{c}{$8.29^{+0.13}_{-0.23}$} & \multicolumn{1}{c}{$0.69^{+0.17}_{-0.10}$} \\
\multicolumn{1}{c}{NGD-z10a$^{*+}$} & \multicolumn{1}{c}{$7.94^{+0.35}_{-0.30}$} & \multicolumn{1}{c}{$0.96^{+0.20}_{-0.19}$} & \multicolumn{1}{c}{$8.15^{+0.21}_{-0.17}$} & \multicolumn{1}{c}{$0.81^{+0.14}_{-0.10}$} & \multicolumn{1}{c}{$7.99^{+0.30}_{-0.35}$} & \multicolumn{1}{c}{$0.90^{+0.25}_{-0.13}$} \\
\multicolumn{1}{c}{NGD-z10b$^{*+}$} & \multicolumn{1}{c}{$8.50^{+0.37}_{-0.09}$} & \multicolumn{1}{c}{$5.93^{+0.26}_{-0.60}$} & \multicolumn{1}{c}{$8.75^{+0.16}_{-0.20}$} & \multicolumn{1}{c}{$4.65^{+0.44}_{-0.30}$} & \multicolumn{1}{c}{$8.70^{+0.17}_{-0.29}$} & \multicolumn{1}{c}{$4.72^{+0.57}_{-0.29}$} \\
\multicolumn{1}{c}{NGD-z10c} & \multicolumn{1}{c}{$8.40^{+0.31}_{-0.11}$} & \multicolumn{1}{c}{$1.67^{+0.16}_{-0.39}$} & \multicolumn{1}{c}{$8.56^{+0.15}_{-0.15}$} & \multicolumn{1}{c}{$1.43^{+0.23}_{-0.15}$} & \multicolumn{1}{c}{$8.56^{+0.15}_{-0.27}$} & \multicolumn{1}{c}{$1.45^{+0.39}_{-0.17}$} \\
\multicolumn{1}{c}{NGD-z10d$^{*+}$} & \multicolumn{1}{c}{$8.15^{+0.35}_{-0.14}$} & \multicolumn{1}{c}{$1.28^{+0.14}_{-0.24}$} & \multicolumn{1}{c}{$8.38^{+0.16}_{-0.21}$} & \multicolumn{1}{c}{$1.06^{+0.20}_{-0.11}$} & \multicolumn{1}{c}{$8.31^{+0.20}_{-0.30}$} & \multicolumn{1}{c}{$1.12^{+0.26}_{-0.13}$} \\
\multicolumn{1}{c}{NGD-z11a$^{*+}$} & \multicolumn{1}{c}{$8.68^{+0.54}_{-0.04}$} & \multicolumn{1}{c}{$5.14^{+1.24}_{-1.66}$} & \multicolumn{1}{c}{$8.99^{+0.23}_{-0.17}$} & \multicolumn{1}{c}{$3.93^{+1.91}_{-0.82}$} & \multicolumn{1}{c}{$8.96^{+0.27}_{-0.32}$} & \multicolumn{1}{c}{$4.61^{+1.69}_{-1.20}$} \\
\multicolumn{1}{c}{NGD-z11b} & \multicolumn{1}{c}{$8.10^{+0.35}_{-0.20}$} & \multicolumn{1}{c}{$1.27^{+0.12}_{-0.28}$} & \multicolumn{1}{c}{$8.33^{+0.16}_{-0.15}$} & \multicolumn{1}{c}{$1.07^{+0.18}_{-0.12}$} & \multicolumn{1}{c}{$8.24^{+0.21}_{-0.34}$} & \multicolumn{1}{c}{$1.12^{+0.26}_{-0.14}$} \\
\multicolumn{1}{c}{NGD-z11c$^{*+}$} & \multicolumn{1}{c}{$8.12^{+0.34}_{-0.14}$} & \multicolumn{1}{c}{$1.37^{+0.09}_{-0.21}$} & \multicolumn{1}{c}{$8.35^{+0.14}_{-0.17}$} & \multicolumn{1}{c}{$1.21^{+0.10}_{-0.10}$} & \multicolumn{1}{c}{$8.25^{+0.21}_{-0.26}$} & \multicolumn{1}{c}{$1.26^{+0.20}_{-0.11}$} \\
\multicolumn{1}{c}{NGD-z12a} & \multicolumn{1}{c}{$8.61^{+0.49}_{-0.11}$} & \multicolumn{1}{c}{$4.47^{+1.17}_{-1.90}$} & \multicolumn{1}{c}{$8.90^{+0.19}_{-0.22}$} & \multicolumn{1}{c}{$3.58^{+1.47}_{-1.22}$} & \multicolumn{1}{c}{$8.86^{+0.25}_{-0.36}$} & \multicolumn{1}{c}{$3.97^{+1.67}_{-1.39}$} \\
\multicolumn{1}{c}{NGD-z15a} & \multicolumn{1}{c}{$8.07^{+0.44}_{-0.11}$} & \multicolumn{1}{c}{$1.79^{+0.24}_{-0.56}$} & \multicolumn{1}{c}{$8.40^{+0.16}_{-0.15}$} & \multicolumn{1}{c}{$1.42^{+0.57}_{-0.23}$} & \multicolumn{1}{c}{$8.32^{+0.18}_{-0.36}$} & \multicolumn{1}{c}{$1.52^{+0.52}_{-0.29}$} \\
\hline
\end{tabular}
\end{table*}

%Add bursts to the SFH as well? No need for this first paper.

%We find that bagpipes often gives far different solutions with different entered priors, with logarithmic priors in metallicity and dust favouring high-z solutions and uniform linear priors favouring the low-z secondary solution present in many of our sample SEDs using NIRCam imaging alone. %To test the effect of this on our sample, we change the form of SFH to analyse the effect of this on our high-z sample.

\subsection{S\'{e}rsic indices and Half-light radii}

\begin{figure}
    \centering
    \includegraphics[width = 0.495\textwidth]{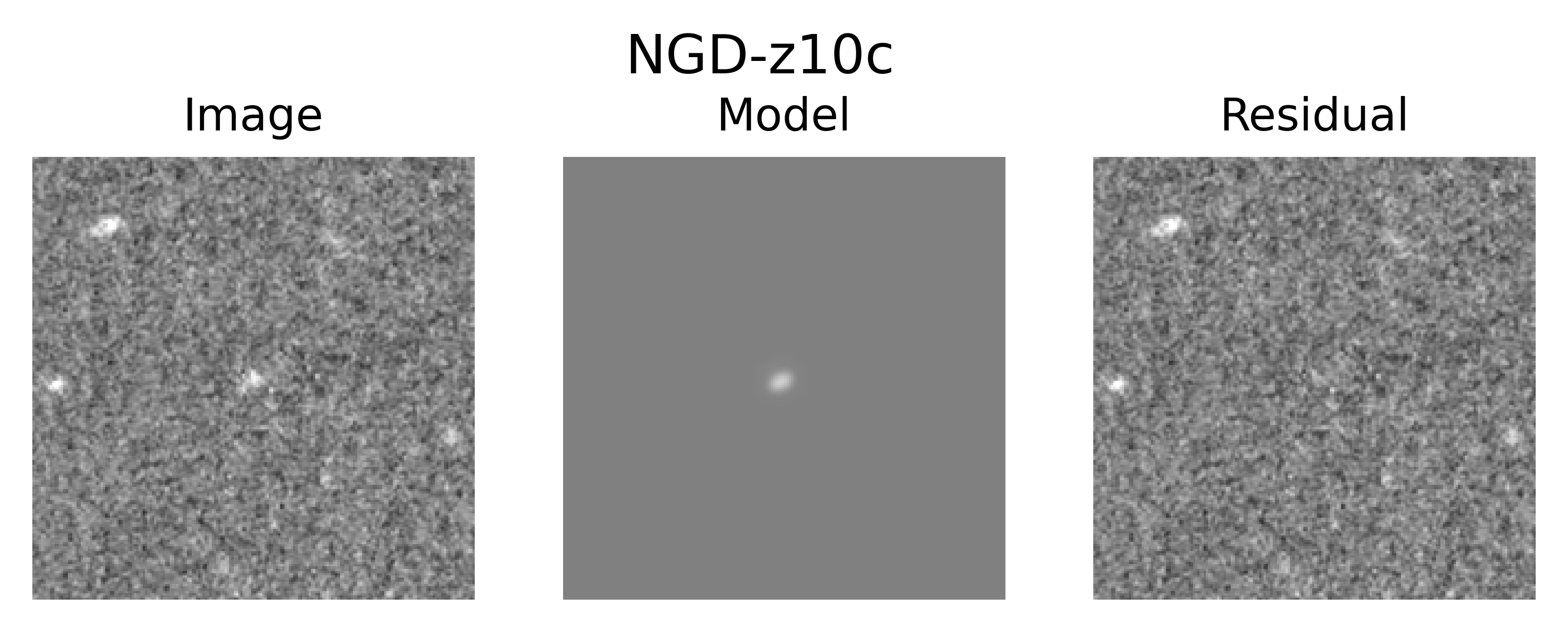}
    \caption{S\'{e}rsic profile fit for NGD-z10c. The left panel shows the source, the central panel shows the two dimensional S\'{e}rsic model, the right panel shows the residual image.}
    \label{fig:morphology}
\end{figure}

We use \galfit\ to determine the S\'{e}rsic indices and half-light radii of these extended sources. \galfit\ is a least-squares fitting algorithm, which uses a Levenberg-Marquardt algorithm to find the optimum solution to a fit \citep{Galfit1, Galfit2}. We follow a similar method to that presented in \citet{Kartaltepe2022}, whereby the \sextractor\ catalogue is used for the initial parameter guesses. We run \galfit\ for all available filters, but only report results for the filters that best match the rest-frame optical wavelength of the source. This minimises the effect of morphological \textit{k}-correction, as the qualitative and quantitative structure of galaxies changes as a function of wavelength \citep{TaylorMager2007}, which can result in significant structural changes between rest-frame UV and rest-frame optical images. In this case, all results reported are in the F444W band.  We fit a 2D S\'{e}rsic profile, with results shown in \autoref{tab:morphology}. An example fit is shown in \autoref{fig:morphology}.

Out of the 18 sources in our sample, 11 of these were flagged due to at least one parameter meeting a constraint limit, or the model being identified as a poor fit by eye, using the residual image created by \galfit\ . The high percentage of poorly fitting models is likely due to the unprecedented depths achieved by \emph{JWST}, resulting in detection of faint sources. It is likely in future that the method of light profile fitting will need some refinement in order to accurately model the faintest sources. 

Most of our modelled galaxies are compact sources ($R_\mathrm{e} < 1$~kpc), with two galaxies being exceptionally small sources with $R_\mathrm{e} < 0.5$~kpc. These small sizes are in agreement with other structural analyses of high redshift objects \citep{Adams2023, compact1}. %The sources with lowest Sersic index are flat galaxies.

\begin{table}
	\centering
	\caption{S\'{e}rsic indices and half light radii for each object for which a clean fit could be obtained using the two dimensional fitting software \galfit. We quote the errors obtained from \galfit\, which are symmetrical errors.}
	\label{tab:morphology}
	\begin{tabular}{ccc}
		Name & S\'{e}rsic Index & $R_\mathrm{e}$ (kpc) \\
		\hline
		NGD-z9a & $1.25\pm0.20$ & $1.34\pm0.23$ \\
        NGD-z8a & $0.79\pm0.25$ & $0.82\pm0.08$ \\
        NGD-z8f & $0.60\pm0.26$ & $0.90\pm0.08$ \\
        NGD-z9c & $2.75\pm2.12$ & $0.37\pm0.11$ \\
        NGD-z10c & $0.07\pm0.85$ & $0.44\pm0.05$ \\
        NGD-z11c & $0.06\pm1.10$ & $0.43\pm0.71$ \\
        NGD-z10d & $0.30\pm0.30$ & $0.98\pm0.17$ \\
        \hline
	\end{tabular}
\end{table}

\section{Discussion}

A source of early debate within the high-redshift community was with regards to the stellar masses of some high redshift candidates \citep[e.g.][]{Labbe2022,Boylin-Kolchin2022,Endsley2022,Lovell2023}. In particular, galaxies with redshifts greater than $z>9$ were found to have stellar masses of up to $10^{11}~\mathrm{M}_{\odot}$. Such high stellar masses raise tension with the $\Lambda$CDM cosmological model \citep{Lovell2023}. The stellar masses obtained in this study are generally much lower and in the regime of $10^{8}~\mathrm{M}_{\odot}$, we compare these to the \citet{Lovell2023} cosmological estimations for the most massive galaxies expected in \emph{JWST}-like survey volumes. Both our highest redshift source and most massive source sit comfortably within the $1\sigma$ and $2\sigma$ regions respectively of these models. Subsequently, the physical properties we measure for these sources show no significant tension with such cosmological models. 

In terms of other simulations,  the stellar masses derived here are in agreement with the results of Astraeus \citep{Hutter2021}, the largest simulation fully coupling galaxy formation and reionization.  This is such that the most massive galaxy we find at $z=9$ seems to hint at the region being slightly over-dense (by about 10\%), as seen from predictions of JWST-JADES surveys accounting for the impact of cosmic variance \citep{Ucci2021}. 

In terms of the number of expected galaxies at our depths and assuming the UV Luminosity Function follows that of recent observations \citep[e.g.][]{Donnan2022,Bouwens2021}, we integrate these UV LF's to predict the number of galaxies that would be expected from NGDEEP at $z=9$ and $z=10$. We find that the number of galaxies identified is broadly consistent with these predictions from past studies (5-8 galaxies at $z=9$ and 1-3 galaxies at $z=10$), indicating that NGDEEP is not extremely over- or under-dense. This also shows that there may not be a large evolution in the luminosity function at fainter absolute magnitudes.  

We also conduct the same simple experiment using the high redshift simulation FLARES, which estimates the form of the UV LF back to $z=10$ using weighted regions of overdense and underdense cosmological volumes \citep{Vijayan2021}. Using the better fit double power-law results, these UV LF's estimate within our area 8 galaxies at $8.5<z<9.5$ and 3 galaxies at $9.5<z<10.5$, identical to the number of sources measured here. A full measure of the UV LF using NGDEEP will be conducted once the image processing is finalised and a full completeness and contamination analysis conducted.

We also compare our derived numbers to results from the Delphi semi-analytic model \citep{Dayal2022} which fully couples the key dust mechanisms (of production, destruction, astration, ejection and grain growth) with galaxy formation at these early epochs. Crucially, this model has been fully baselined against the latest Atacama Large millimeter Array (ALMA) dust estimates at $z \sim 7$ \citep{bouwens2022}. Accounting for the redshift-dependent impact of dust attenuation, this model predicts 12 and 4 galaxies brighter than a magnitude of 29.5 at $z \geq 10$ and 12, respectively. These numbers are in excellent agreement with our derived values in this work.  Thus, unlike in some other studies, we do not see an obvious problem within this deep, but small, field in terms of comparisons to CDM based cosmological models. 

In terms of star formation rates amongst our sample, NGD-z9a shows a much higher SFR ($209.18~\mathrm{M}_{\odot}\mathrm{yr}^{-1}$) than other galaxies in our sample. One explanation for this high SFR is that it could include an AGN contribution. Further sub-mm observations (e.g. ALMA) could provide far-infrared information to make this clear. In addition, we believe that the neighbouring bright source may be increasing its Kron radius, meaning that we are likely over-correcting using the currently implemented FLUX\_AUTO to aperture corrected FLUX\_APER ratio, which will boost both the mass and SFR of this source. The scale of this correction is currently half a dex.

\section{Summary and Conclusions}
\label{sec:conclusions}

  We provide in this paper a first view and analysis of the galaxies found within the NGDEEP field, which will be the deepest public NIRCam imaging set once complete.   We find 18 $8 < z < 16$ galaxies through our bespoke reduction, analysis and selection methods identified using two different photometric redshift codes; \lephare\ and \eazy.

Even though incomplete in terms of its depth, the NGDEEP survey has allowed us to identify galaxies with low inferred stellar mass.  These galaxies provide a new window into the early Universe, allowing us to study the formation and evolution of galaxies as well as increase our understanding of the faintest and most distant objects in the Universe.   

 Our major conclusions can be summarized as follows. We find a significant number of low mass dwarf galaxies with $M_{*} < 10^{8.5}$~\solm.   One of our objects is at $z=15.57^{+0.39}_{-0.38}$, which we find has a blue UV slope of 
$\beta=-3.25^{+0.41}_{-0.46}$ and a stellar mass of $M_{*} = 10^{7.39}$~\solm.   The majority of the galaxies in this field are faint, low-mass galaxies, at $z \sim 9$ that have blue colors and UV slopes.  We find in general that these galaxies all have flat surface brightness profiles and are small with $R_\mathrm{e}< 1$~kpc. 

Thus our major finding is that there is a significant population of low mass and low luminosity galaxies in the epoch of reionization that we are just now beginning to be able to study with \emph{JWST}.  Future studies will determine the contribution of these populations to the UV flux at high redshift and its contribution to reionization.

The discovery of these distant galaxies is a major step forward in our understanding of the Universe and its history, and the information gathered by deep NIRCam  \emph{JWST} will be crucial in furthering our knowledge of the early Universe and the processes that shaped it.

\section*{Acknowledgements}

We thank the NGDEEP team for their work in designing and preparing these observations, the STScI staff that carried them out and the NIRISS team, whose prompt action to resolve technical issues at the end of January 2023 enabled this data to be taken.

We acknowledge support from the ERC Advanced Investigator Grant EPOCHS (788113; PI Conselice), as well as two studentships from STFC to DA and TH. LF acknowledges financial support from Coordenação de Aperfeiçoamento de Pessoal de Nível Superior - Brazil (CAPES) in the form of a PhD studentship.  This work is based on observations made with the NASA/ESA \textit{Hubble Space Telescope} (\emph{HST}) and NASA/ESA/CSA \textit{James Webb Space Telescope} (\emph{JWST}) obtained from the \texttt{Mikulski Archive for Space Telescopes} (\texttt{MAST}) at the \textit{Space Telescope Science Institute} (STScI), which is operated by the Association of Universities for Research in Astronomy, Inc., under NASA contract NAS 5-03127 for \emph{JWST}, and NAS 5–26555 for \emph{HST}. The DOI connecting this publication to the raw data source is \url{http://dx.doi.org/10.17909/v7ke-ze45}. PD acknowledges support from the NWO grant 016.VIDI.189.162 (``ODIN") and the European Commission's and University of Groningen's CO-FUND Rosalind Franklin program.

\bibliography{main}{}
\bibliographystyle{aasjournal}



\end{document}